\documentclass[aps, superscriptaddress, prx,11pt,longbibliography,nofootinbib,tightenlines]{revtex4}
\usepackage{graphicx,hyperref}
\usepackage{amsmath}
\usepackage{amsfonts}
\usepackage{epsfig} 
\usepackage{color}
\usepackage{bbm}
\usepackage{latexsym} 
\usepackage{amsmath}
\usepackage{amssymb}
\usepackage{amsthm}

\newcommand{\beq}{\begin{equation}}
\newcommand{\eeq}{\end{equation}}
\newcommand{\bea}{\begin{eqnarray}}
\newcommand{\eea}{\end{eqnarray}}

\mathchardef\nss="711B

\def\ket#1{{\left|#1\right\rangle}}
\def\bra#1{{\left\langle #1 \right|}}

\def\nss{G}

\def\be{\begin{eqnarray}}
\def\ee{\end{eqnarray}}

\newlength{\myL}

\newcommand{\<}{\langle}

\newcommand{\up}{\uparrow}
\newcommand{\down}{\downarrow}
\renewcommand{\>}{\rangle}
\renewcommand{\(}{\left(}
\renewcommand{\)}{\right)}
\renewcommand{\[}{\left[}
\renewcommand{\]}{\right]}

\renewcommand{\d}{\partial}

\newcommand{\avg}{\mathop{\mathbb{E}}}

\newcommand{\dia}[3]{\raisebox{#3pt}{\includegraphics[height=#2pt]{dia_#1}}}

\makeatletter
\newcommand\xleftrightarrow[2][]{%
  \ext@arrow 9999{\longleftrightarrowfill@}{#1}{#2}}
\newcommand\longleftrightarrowfill@{%
  \arrowfill@\leftarrow\relbar\rightarrow}
\makeatother

\begin{document}

\title{Entanglement dynamics in hybrid quantum circuits}

\author{Andrew C. Potter}
\affiliation{Department of Physics and Astronomy, and Quantum Matter Institute,
University of British Columbia, Vancouver, BC, Canada V6T 1Z1}

\author{Romain Vasseur}
\affiliation{Department of Physics, University of Massachusetts, Amherst, MA 01003, USA}

\date{\today}

\begin{abstract}
The central philosophy of statistical mechanics (stat-mech) and random-matrix theory of complex systems is that while individual instances are essentially intractable to simulate, the statistical properties of random ensembles obey simple universal ``laws".
This same philosophy promises powerful methods for studying the dynamics of quantum information in ideal and noisy quantum circuits --  for which classical description of individual circuits is expected to be generically intractable. 
Here, we review recent progress in understanding the dynamics of quantum information in ensembles of random quantum circuits, through a stat-mech lens. 
We begin by reviewing discoveries of universal features of entanglement growth, operator spreading, thermalization, and chaos in unitary random quantum circuits, and their relation to stat-mech problems of random surface growth and noisy hydrodynamics.  
We then explore the dynamics of monitored random circuits, which can loosely be thought of as noisy dynamics arising from an environment monitoring the system, and exhibit new types of measurement-induced phases and criticality. 
Throughout, we attempt to give a pedagogical introduction to various technical methods, and to highlight emerging connections between concepts in stat-mech, quantum information and quantum communication theory.
\end{abstract}

\maketitle

\tableofcontents

\section{Introduction}
While many-body physics has traditionally focused on the properties of cold-matter in equilibrium, emerging atomic, molecular, optical, and qubit platforms allow access to far from equilibrium dynamics with local space and time control over interactions. A key challenge is to identify universal features of non-equilibrium quantum dynamics and approach to thermalization. The dynamics of the scrambling of local quantum information into non-local degrees of freedom  by many-body unitary dynamics plays a central role in addressing those questions. The growth of entanglement is not only important to diagnose thermalization (or lack thereof), but also characterizes the complexity of tensor network descriptions of quantum dynamics. Of particular interest are {\em universal} properties, that do not depend on particular microscopic details, and hold for generic quantum systems.

Insights into generic non-equilibrium  dynamics can be gained by considering minimally structured models, such as {\em random unitary circuits}~\cite{HaydenPreskill2007,2008Fastscramblers,Brown:2015aa,Hosur:2016aa,PhysRevLett.98.130502,Brandao:2016aa,PhysRevX.7.031016,Nahum2018}, which capture the salient ingredients of generic quantum systems, namely {\em unitarity} of the dynamics, and {\em locality} of the interactions. Using random quantum circuits, the growth of entanglement in one dimensional system was elegantly mapped to the celebrated Kardar-Parisi-Zhang universality class~\cite{PhysRevLett.56.889} of random surface growth~\cite{PhysRevX.7.031016}. This mapping also uncovered deep connections to the Ryu-Takayanagi formula in in holographic AdS/CFT correspondences~\cite{PhysRevLett.96.181602,RT2006}, by establishing a relation between  entanglement dynamics and a geometric minimal-cut picture. Random unitary circuits were also used to characterize exactly the local spreading of operators in the Heinsenberg picture~\cite{Nahum2018,VonKeyserlingk2018}, providing a complementary picture on chaotic dynamics and scrambling from the perspective of operator growth. Other probes of many-body quantum chaos, {\it e.g.} related to level statistics, have also been computed exactly in Floquet (time-periodic) circuits~\cite{PhysRevX.8.041019,PhysRevLett.121.264101,PhysRevLett.121.060601,Friedman2019,PhysRevB.100.064309,PhysRevLett.123.210601}. In turn, those exact results led to a coarse-grained, ``hydrodynamic'' description of operator spreading that was conjectured to universally apply to non-integrable quantum systems in one dimension. Since then, random quantum circuits have become part of the standard toolbox to study chaotic quantum dynamics, and provided crucial insights into, {\it e.g} the emergence of irreversible hydrodynamics from unitary evolution in the presence of a conserved charge~\cite{PhysRevX.8.031057,PhysRevX.8.031058}. 

Motivated by the advent of noisy intermediate-scale quantum simulators~\cite{Preskill2018quantumcomputingin}, random quantum circuits have also been used to study the dynamics of entanglement in open quantum systems, which are continuously ``monitored'' by their environments. Non-unitary random circuits provide a natural tool to study the competition between unitary dynamics, which leads to chaotic evolution and rapid entanglement growth, and non-unitary operations resulting from measurements and noisy couplings to the environment, that tend to irreversibly destroy quantum information by revealing it~\cite{PhysRevB.98.205136,Skinner2019}. More generally, non-unitary circuits and random tensor networks~\cite{RTN,PhysRevB.100.134203,PhysRevB.102.064202,Nahum2020,2021arXiv210802225L,2021arXiv210712376Y} can exhibit a variety of ``phases'' and phase transitions with different entanglement scalings. The most studied representative example of such an entanglement transition that results from the competition between unitary dynamics and non-unitary processes is the so-called measurement-induced phase transition~\cite{PhysRevB.98.205136,Skinner2019}. This transition occurs in  monitored random quantum circuits (MRCs) made up of random unitary gates, combined with local projective measurements occurring at a fixed rate, separating two phases with very different entanglement properties. Importantly, such measurement-induced phase transitions (MIPTs) are only visible in an individual quantum trajectory (i.e., the pure state of the system conditional on a set of measurement outcomes), and in trajectory-averages of quantities that are nonlinear functions of the density matrix. When measurements are frequent enough, they are able to efficiently extract quantum information from any initial state, and Zeno collapse it into a weakly-entangled state with area law entanglement. In contrast, for a small enough measurement rate, the unitary dynamics scrambles quantum information into nonlocal degrees of freedom that can partly evade local measurements. In this entangling phase (volume law), initial product states become highly entangled over time, while initial mixed states remain mixed for extremely long times~\cite{Gullans2019}. There are different perspectives on this measurement-induced transition, either as a purification transition~\cite{Gullans2019} or from the language of quantum communication and error correction. In the volume law phase, the unitary dynamics is effectively able to hide non-local degrees of freedom that span a decoherence-free subspace in which the dynamics is effectively unitary~\cite{Choi2020,Gullans2019,Li2020b,fan2020self}: this subspace can be regarded as the code space of a quantum error correcting code.

Measurement-induced transitions have been investigated numerically and theoretically in various contexts, dimensionality, geometries, with different families of gates~\cite{PhysRevB.98.205136,Skinner2019,Li2019,PhysRevB.99.224307,Li2020,10.21468/SciPostPhys.7.2.024,Choi2020,Gullans2019,Szyniszewski2019,Bao2020,Jian2020,Gullans2020,Zabalo2020,PhysRevResearch.2.023288,Ippoliti2020,Lavasani2020,Sang2020,PhysRevResearch.2.013022,PhysRevB.102.064202,Nahum2020,Turkeshi2020,Fuji2020,Lunt, Lunt2020,fan2020self,2020arXiv200503052V,Li2020b,PhysRevB.103.224210,PhysRevLett.126.060501,2021arXiv210306356L,2020arXiv201204666J,PhysRevLett.126.170503,2021arXiv210106245T,2021arXiv210209164B,2021arXiv210413372B,2021arXiv210407688B,2021arXiv210703393Z,2021arXiv210710279A,2021arXiv210804274L,PhysRevLett.126.170602,2021arXiv210609635J,2021arXiv210410451D,2021arXiv210906882S,2020arXiv201203857L,2021arXiv210208381B,sierant2021dissipative,2021arXiv211014403S}, establishing that it is a generic property of quantum trajectories of open quantum systems. A particularly fruitful approach to understand the phenomenology and universality class of measurement-induced transitions is to use exact mappings onto effective statistical mechanics models, that emerge from using a replica trick to deal with the intrinsic non-linearities of the problem, and after averaging over the random gates.
This systematic  statistical mechanics approach based on replica permutation ``spins" was first developed in the context of random tensor networks~\cite{RTN,PhysRevB.100.134203}, and then extended to deal with random unitary~\cite{Zhou2019} and monitored~\cite{Bao2020,Jian2020} circuit dynamics.
 Such stat-mech mappings provide an appealing picture of the entanglement transition in terms of a (replica) symmetry-breaking transition, where the volume-law coefficient of entanglement has a simple interpretation as a domain wall surface tension. In turn, these recent theoretical developments raise the intriguing prospect of using well-developed statistical mechanics tools to study quantum communication channel capacity and error-correction thresholds~\cite{Choi2020,Gullans2019,fan2020self,Li2020b,PhysRevX.11.031066,2021arXiv210804274L}, and computational complexity~\cite{2019arXiv190512053H,Napp2019}. Finite-size evidence for such an entanglement MIPT was even recently observed experimentally in trapped-ion chains~\cite{noel2021}. The phase structure and dynamics of non-unitary circuits is being actively explored at the time of writing of this chapter. 

The outline of this chapter is as follows: in Section.~\ref{secRandomCircuits}, we introduce random quantum circuit dynamics, and derive exact results on entanglement growth and operator spreading in such circuits. We also comment on the role of symmetries in the dynamics. In Section~\ref{secMeasurements}, we introduce measurements, and describe the phenomenology of measurement-induced entanglement transitions from different perspectives. Section~\ref{secStatMech} derives exact statistical mechanics mappings for random unitary circuits with and without measurements, using a replica trick. Various consequences for entanglement dynamics and criticality are discussed. Finally, in Section~\ref{SecSymmetryTopology}, we discuss progress in understanding measurement-induced symmetry-breaking and topological orders and related criticality, which would be forbidden in equilibrium and are stabilized by dissipation.

\section{Random unitary quantum circuits}
\label{secRandomCircuits}

We begin by studying entanglement growth in ensembles of unitary random quantum circuit (RC) dynamics. The growth of entanglement is an important metric to diagnose dynamical thermalization and distinguish this behavior from other dynamical universality classes such as many-body localization. Entanglement growth also reflects the complexity of matrix-product state (MPS) and tensor-network state (TNS) descriptions of quantum dynamics. 
Just as in the statistical mechanics of many interacting particles, or random matrix theory of complex Hamiltonians, the statistical properties of random ensembles of circuits can often be captured with a far-simpler theoretical description than the (generically exponentially-difficult) task of computing the detailed output of a given circuit instance.
In this section we review two equivalently complementary perspectives of entanglement growth and thermalization in RCs, first we work in the ``Schr\"odinger" picture and examine the growth of bipartite entanglement in the evolution of quantum states, and second we adopt a ``Heisenberg" picture and examine the evolution and spreading of operators under RC dynamics. These two descriptions are elegantly united~\cite{PhysRevX.7.031016} in a mapping of the entanglement growth problem onto Kardar-Parisi-Zhang (KPZ) dynamics of random surface growth~\cite{PhysRevLett.56.889}. We recount the connection of surface growth to an equivalent picture in terms of directed random polymers, which has a geometrical interpretation closely analogous to the Ryu-Takanagi relation between geometry and entanglement in holographic AdS/CFT correspondences~\cite{PhysRevLett.96.181602,RT2006}. 

\subsection{Entanglement growth}
For this discussion, we adopt the model of~\cite{PhysRevX.7.031016}, which consists of a length $L$ chain of $d$-level qudits, in which, at every time-step a randomly selected pair of neighboring qubits is subjected to a random two-qudit entangling gate drawn uniformly from the Haar distribution on U$(d^2)$. To study entanglement growth, consider the bipartite von Neumann entropy: $S(x,t) = -\text{tr}\rho_{[x+1,\infty)}\log_d \rho_{[x+1,\infty)}$, for bipartitioning the system across the bond $(x,x+1)$, where $\rho_{[x,x']}$ is the reduced density matrices for sites in the interval $[x,x']$. 

\subsubsection{Mapping to KPZ dynamics of random surface growth}
Due to the subadditivity property of entanglement, the entanglement entropy for neighboring cuts is bounded by: 
\begin{align}
 |S(x,t) - S(x+1,t)| \leq 1.
\end{align}
i.e. if $S$ behaves like a ``surface" with bounded slope. The stochastic dynamics of this ``surface" can be understood by a simple heuristic rule. When a gate acts on bond $(x,x+1)$ the corresponding entanglement grows as:
\begin{align}
S(x,t+1) = \text{min}\{S(x-1,t),S(x+1,t)\}+1,
\label{eq:Supdate}
\end{align}
which turns out to be exact (almost surely) in the limit of large onsite dimension $d\rightarrow \infty$, but which is believed to capture the universal aspects of RC dynamics for any finite-$d$. This rule can be understood as follows: if $S(x,t)=S(x-1,t)$ or $S(x-1,t)-1$, i.e. implying that site $x+1$ is unentangled with $(-\infty,x]$ at time $t$, then the gate increases the entanglement by an amount that is generically proportional to $1$, and becomes precisely $1$ (almost surely)  in the large-$d$ limit. If on the other hand, $S(x,t)= S(x-1,t)+1$, this implies that site $x+1$ is already maximally entangled with $(-\infty,x]$ and the gate is very unlikely to disentangle it, so $S(x,t)$ remains unchanged. We can see that since $S(x,0)=0~\forall x$, and since $S(x,t)$ changes by quantized amounts (at $d\rightarrow \infty$), then $S(x,t)$ is always an integer, and these cases exhaust the possibilities. 
Concrete examples of updates are shown in Fig.~\ref{fig:mincut}a,b. Notice that locally flat parts of the surface tend to grow (top left of Fig.~\ref{fig:mincut}b, while bonds with negative local curvature get converted to having local positive curvature (bottom left of Fig.~\ref{fig:mincut}b), and regions with positive slope of $s$ do not grow (bottom right of Fig.~\ref{fig:mincut}b).

To study the long-time dynamics it is useful to ``zoom-out" and coarse-grain the spatial lattice and integral step of the circuit by introducing an average entropy $s(x,t)$ over blocks of sites of size $\ell\gg 1$ (similar to moving from a lattice description of magnetic moments, to a continuum description of coarse-grained average magnetization), and coarse grain our time step by $L$ so that a finite-density of gates are applied in a coarse-grained time-step. The continuum limit of the ``update rule" Eq.~\ref{eq:Supdate} can then be written in the form of a KPZ equation:
\begin{align}
\frac{\d s}{\d t} = \nu \d_x^2 s - \frac\lambda 2 (\d_x s)^2 +\eta(x,t)+c,
\label{eq:kpz}
\end{align}
where $c$ reflects the average rate of growth of entanglement, $\eta(x,t)$ is a random noise (capturing the stochastic placement of gates in space and time in the RC model), the $\nu$ terms suppresses local curvature reflecting suppression of negative curvature from processes like the one shown in the bottom left of Fig.~\ref{fig:mincut}b, and finally the $\lambda$ term reflects that the entanglement growth is slower in regions with non-zero slope of $s$ as described above. Other terms with more derivatives or higher nonlinearities are irrelevant in the renormalization group sense, so that this KPZ equation captures the universal aspects of the coarse-grained entanglement growth.

The overall trend of Eq.~\ref{eq:kpz} is that entanglement grows linearly in time $\<s(x,t)\>\sim v_Et+\dots$ with constant ``entanglement-velocity" $v_E$, and $(\dots)$ reflects subleading contributions that grow more slowly than $\sim t$. Universal scaling of fluctuations about this average trend are governed by the exactly-solved KPZ universality class. Measurable quantities at distance $x$ and time $t$ scale as universal functions of the ratio $x/\xi(t)$ with correlation length $\xi(t)\sim t^{1/z}$ where the dynamical exponent $z=\frac32$. For example, the difference in entanglement for cuts separated by distance $r$ at equal times $t$ scales like  $\<\(S(x+r,t)-S(x,t)\)^2\>\sim r^\alpha g\(r/\xi(t)\)$ where $g$ is a universal function and $\alpha=\frac12$. Another important critical exponent, $\beta$, characterizes the RMS fluctuations in entanglement: $\sqrt{\<(s(x,t))^2\>-\<s(x,t)\>^2}\sim t^\beta$ with $\beta=\frac13$, and similarly controls the dominant sub-leading correction to the entanglement growth: $\<s(x,t)\>\sim v_Et+Bt^\beta$.

\subsubsection{Directed polymer and minimal cut interpretation}
\begin{figure}[bt!] 
    \centering
    \includegraphics[width=0.75\textwidth]{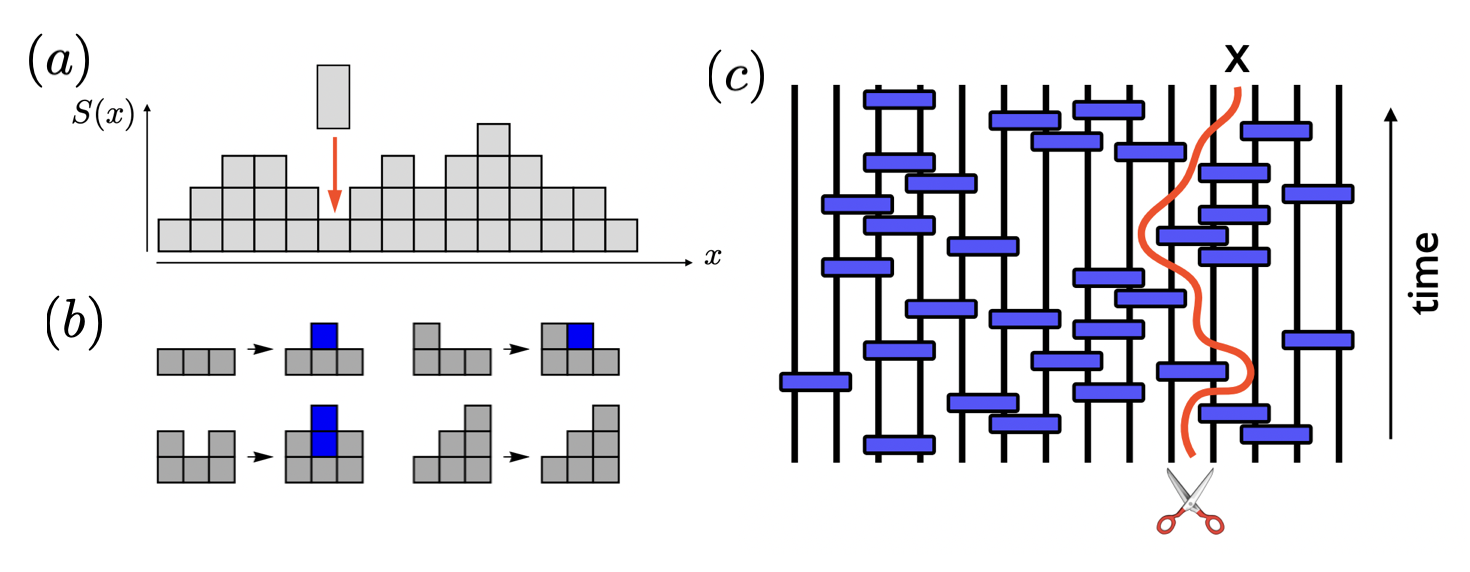}
    \caption{ {\bf KPZ dynamics of entanglement growth in random circuits} In a random circuit model consisting of randomly placed nearest-neighbor two-site gates (c), the entanglement dynamics as a random surface growth where each gate on a bond tends to increase the height at that bond by one block (a). (b) shows the surface-growth steps for a gate acting on a bond with different pre-existing entanglement configurations. (c) Shows an alternative, but equivalent picture of entanglement growth in terms of a minimal cut: entanglement is upper-bounded by the minimal number of bonds cut by a line that bipartitions the circuit without cutting through any gates. This minimal cut can be viewed as the minimal-free-energy configuration of a directed polymer in a random medium, which is well-known to be equivalent to the random surface growth model. Reproduced from Ref.~\cite{PhysRevX.7.031016}.
    \label{fig:mincut} 
            }
\end{figure}

KPZ dynamics arises in a wide variety of problems besides random surface growth. A prominent example is the dynamics of a directed polymer in a random environment, i.e. a sequence of monomer segments arranged along one direction (``time") which can be incline but not turn back on itself. This directed-polymer perspective on KPZ dynamics has a natural geometric interpretation in the RC dynamics.
To estimate $S(x,t)$, consider the following geometric construction: draw a curve through the random circuit starting from bond $(x,x+1)$ at time $t$ and moving back through the circuit to time $t=0$ without crossing any gates. The length of this upper-bounds the Schmidt rank of the bipartition across bond $(x,x+1)$ (see Fig.~\ref{fig:mincut}), and hence also upper-bounds the entropy $S(x,t)$. The best upper-bound estimate for $S(x,t)$ from this procedure is therefore given by the length of the minimal such cut through the circuit. Ref.~\onlinecite{PhysRevX.7.031016} showed that this upper-bound is in fact saturated in the large-$d$ limit.
Identifying the cut-line with the shape of a directed polymer, and the constraints imposed by the constraint that the line does not cross the (randomly placed) gates as a random environment, then the problem of finding minimum length cut is equivalent to minimizing the free energy of this directed polymer in said random environment. This picture led to a more general and universal ``entanglement membrane'' formalism to compute entanglement in chaotic quantum systems~\cite{PhysRevX.7.031016,2018arXiv180300089J,PhysRevX.10.031066}. 

This geometric interpretation of entanglement of a quantum state output by a RC via the geometry of a minimal surface through the circuit is strikingly reminiscent of the Ryu-Takanagi formula  relating the entanglement of a spatial region of a conformal field theory ground-state to the surface area of a spanning geodesic the dual bulk gravity description~\cite{RT2006,PhysRevLett.96.181602}, and to similar relation in high-bond dimension tensor networks which can in a sense be viewed as lattice discretizations of the gravity/CFT correspondence~\cite{RTN}.

\subsection{Operator spreading}
Examining the evolution of operators under RC dynamics provides a complementary perspective on thermalization and chaos. The spreading of operators in the Heisenberg picture can be computed exactly for random quantum circuits~\cite{Nahum2018,VonKeyserlingk2018}. Let us focus on the case of $d=2$ (qubits) for simplicity, although the concepts will extend straightforwardly to arbitrary $d$, and consider the dynamics of an initially local operator ${\cal O}$, which we take to be a Pauli matrix. Under unitary time evolution, this operator is going to become more complicated, and we expand it onto the basis of {\em Pauli strings} ${\cal S}$
\begin{equation}
{\cal O}(t) = U^\dagger(t) {\cal O}(0) U(t)= \sum_{{\cal S}} a_{\cal S}(t) {\cal S},
\end{equation}
where ${\cal S}$ are any product of Pauli matrices on distinct sites. Unitarity as well as the normalization of the initial operator implies the conservation law: 
$\sum_{{\cal S}} \left| a_{\cal S}(t) \right|^2 =\sum_{{\cal S}} \left| a_{\cal S}(0) \right|^2\equiv 1$. The average evolution of the weights $\left| a_{\cal S} \right|^2$ under random Haar evolution is particularly easy to work out. Let us consider a single gate, acting on two sites for simplicity. There are two cases: if the string ${\cal S} $ happens to be the identity $I$ at time $t$, then it is unchanged by the unitary gate and remains identity at time $t+1$. On the other hand, if the string ${\cal S} $  is any other operator in the operator Hilbert space, the random Haar evolution evolves it to any other non-identity string with equal probability. This means that the string weights obey the Markov process:
\begin{equation} \label{eqHaarOperatorWeights}
\left| a_{\cal S} (t+1) \right|^2 = \sum_{\cal S'} {\cal W}_{\cal S, \cal S'} \left| a_{\cal S'} (t) \right|^2,
\end{equation}
with the Markov operator
\begin{equation}
{\cal W}_{\cal S, \cal S'} = \delta_{\cal S, I} \delta_{\cal S',I} + \frac{(1- \delta_{\cal S, I}) (1- \delta_{\cal S', I})}{d^4-1}.
\end{equation}
Instead of keeping track of the $d^L$ coefficients $a_{\cal S}$, it is convenient to focus on simpler quantities. Here we follow Refs~\onlinecite{Nahum2018,VonKeyserlingk2018}, and focus on the {\em right-weight} $\rho(x,t)$, defined as the fraction of strings ending at position $x$
\begin{equation}
\rho(x,t) = \sum_{{\cal S} \textrm{ ending at } x} \left| a_{\cal S} (t) \right|^2.
\end{equation}
Intuitively, the right weight keeps track of the ``operator front'', and can also be related to out-of-time ordered correlators~\cite{Nahum2018,VonKeyserlingk2018} that are used in diagnostics of quantum chaos. Because the right-weight is locally conserved, $\sum_x \rho(x,t)=1$, we expect it to obey a coarse-grained (hydrodynamic) continuity equation 
\begin{equation} \label{eqHydroRightWeight}
\partial_t \rho + \partial_x j=0.
\end{equation}
The dynamics of the right-weight can be readily understood from the Markov process~\eqref{eqHaarOperatorWeights}. Consider the action of a random unitary gate acting on sites $x$ and $x+1$, on a string at ending at position $x$ at time $t$. There are only $d^2-1$ operators out of $d^4-1$ non-identity operators that the random evolution can generate with an identity operator at position $x$. We conclude that with probability $p = (d^2-1)/(d^4-1)$, the operator front does not move and remains at position $x$ (on average), whereas it moves right with probability $1-p$. This immediately implies that the right-weight follows a {\em biased random walk}: the operator front moves ballistically to the right since $p<1/2$, and will broaden diffusively with time as $\sim \sqrt{t}$. Using standard results, this implies that the current in the hydrodynamic equation~\eqref{eqHydroRightWeight} can be expressed within a gradient expansion as
\begin{equation} 
j = v_B \rho - D \partial_x \rho + \dots,
\end{equation}
with $v_B = 1-2p = \frac{d^2-1}{d^2+1}$, and $D = 2 p (1-p) = \frac{2d^2}{(1+d^2)^2}$. This right-weight thus behaves as $\rho(x,t) \sim \frac{1}{\sqrt{t} } {\rm e}^{-(x-v_B t)^2/(4 D t)}$ at long times, and as $d\to \infty$, the front becomes sharp since $D\to 0$, and $v_B \to 1$. This diffusive broadening of the operator front is believed to be a generic feature of chaotic (non-integrable) quantum systems in one dimension. (The operator front also broadens diffusively in interacting integrable systems through a very different physical mechanism~\cite{PhysRevB.98.220303}.) Note that one-dimensional hydrodynamics is known to be unstable in one dimension~\cite{PhysRevA.16.732}: sound waves that would naively broaden diffusively acquire some anomalous KPZ scaling in fluctuating hydrodynamics due to non-linearities. This does not happen in the context of operator spreading as the butterfly velocity $v_B$ does not depend linearly on the right-weight, and the operator front is believed to generically broaden diffusively rather than with KPZ dynamics. 

\subsection{$U(1)$ symmetric circuits} \label{SectionU1}

Random quantum circuits can be enriched by including global symmetries~\cite{PhysRevX.8.031058,PhysRevX.8.031057}, for example adding a conserved $Q = \sum_x q(x)$, and demanding that random gates decompose into Haar random operations within each block of total charge in order to ensure charge conservation. There are several ways to enforce charge conservation: following~\cite{PhysRevX.8.031057}, we can consider a one-dimensional chain in which each site hosts a two-level system (``qubit'') whose computational basis states $\{|0\>,|1\>\}$ have charge $q=0,1$ respectively, and an auxiliary $d$-level system (``qudit'') of charge-neutral degrees of freedom, i.e., with on-site Hilbert space $\mathbb{C}^2 \otimes \mathbb{C}^d$.
The dynamics will consist of local unitary gates and measurements, which are chosen to conserve the U(1) charge associated with the $z$ component of the qubits. 
As before, the symmetry-preserving two-site unitary gates are arranged in a brickwork geometry, but now take the form: 
\begin{align}
U_{i,i+1} = \begin{pmatrix}
U^0_{d^2\times d^2} &0 & 0  \\
0 & U^1_{2d^2\times 2d^2} &0 \\
0 & 0 & U^2_{d^2 \times d^2}
\end{pmatrix}, \label{eq: Unitary matrix structure}
\end{align}
where $i$ labels a site, $U^{q}_{D\times D}$ is a unitary matrix of size $D\times D$ acting on the charge $q_1 + q_2 = q \in \{0,1,2\}$ sector (a local charge is defined to take values $0$ and $1$), and $D$ is the dimension of the Hilbert space of the charge sector. 
Each matrix is drawn independently from the Haar random ensemble of unitary matrices of the appropriate size.

Those random unitary gates spread the charge uniformly with equal probability, so the charge performs a random walk
\begin{equation} 
q(x,t+1) = q(x+1,t+1) = \frac{1}{2} \left( q(x,t)+  q(x+1,t) \right),
\end{equation}
independently of the onsite Hilbert space dimension $d$. Upon coarse-graining, this naturally leads to a diffusion equation for the local charge
\begin{equation} 
\partial_t q = D_q \partial^2_x q,
\end{equation}
with the diffusion constant $D_q=1/2$ for all $d$. Charge diffusion has some interesting consequences on operator spreading that we will not discuss here, we refer the reader to the original references~\onlinecite{PhysRevX.8.031058,PhysRevX.8.031057}.

Charge conservation has particularly dramatic consequences on the dynamics of entanglement~\cite{Rakovszky2019,Huang2019,Zhou2020,Rakovszky2020,Znidaric2020}: the charge contributions to the Renyi entropies grow diffusively, $S_{n>1} \sim \sqrt{t}$, while the von-Neumann entropy remains ballistic as in the absence of symmetry $S_{n=1} \sim t$. This phenomenon arises from rare fluctuations that leave a region empty (or maximally filled). Consider for concreteness $d=1$ (no neutral degree of freedom), corresponding to the onsite charge states $q=0$ and $q=1$. We are interested in the entanglement across a cut at $L/2$ following the dynamics of an initial product state for the qubit such as $|\psi\rangle = \otimes_{i = 1}^L |+\rangle_i$ where $|+\rangle = \frac{1}{\sqrt{2}} (|0\rangle + |1\rangle)$, (the generalization to other initial states will be readily apparent). We can divide the system into three regions: a central region of radius $\ell = \sqrt{D t}$ centered at the entanglement cut, and regions to the left and right. We then define a configuration to have a ``dead-region" of size $\ell$ if the spins in a region of size $\sim \ell$ are either all $0$, or all $1$, e.g. $|\psi_{\text{dead}} \rangle = \left| \dots+++00\dots 00+++ \dots \right. \rangle$. 
The amplitude for this state in the initial configuration is generically exponentially small in $\ell$ (e.g. for the particular initial and dead-region states mentioned above it is $\sim 2^{-\ell/2}$). So one might be tempted to ignore contributions from large dead regions with $\ell \gg 1$. 
However, a crucial point is that these rare dead regions make an outsized contribution to Renyi entropies with Renyi index $n>1$. To see this, consider the evolution for time $t$ of an amplitude in the initial state having a dead region of size $\ell \gtrsim t^2/D_q$. In this time, particles begin to diffuse into the dead region from the edge, but do not have time to fluctuate across the entanglement cut. Hence, the time-evolved state is still separable into left- and right- Schmidt states. 
The Schmidt weight for such rare dead-regions is given by their probability to occur in the initial state, which is $\approx 2^{-\ell}\approx 2^{-\sqrt{Dt}}$. By contrast, typical configurations without dead-regions all evolve into highly-entangled states with much smaller Schmidt weight $\approx 2^{-v_Et}$, where $v_E$ is the entanglement velocity.
%
%
All R\'enyi entropies with $n > 1$ are dominated by the log of the largest Schmidt coefficient and grow as $\sqrt{t}$. The Von Neumann entropy $S_1$ is dominated instead by \emph{typical} Schmidt coefficients: the number of these grows exponentially in $t$, but they are also exponentially small in $t$ and are therefore subleading for $n > 1$.
We note that, in systems with both charged- and neutral- degrees of freedom (such as the qubit$\times$qudit model mentioned above), this diffusive charge-contributions to Renyi entropy add to a dominant ballistic ($\sim t$) growth from neutral degrees of freedom. However, in a purely-charged models with bounded maximal or minimal charge on each site, Renyi entropies will always grow diffusively.

\section{Measurement-induced phase transitions} \label{secMeasurements}
The random-circuit dynamics discussed above represents the unitary evolution of an ideal closed quantum system. In practice, no system is truly isolated and understanding the interplay of unitary operations with environmental noise and decoherence is a key challenge for quantum computing. Environmental decoherence can be modeled as the environment ``monitoring" (i.e. effectively measuring) the system, which we will idealize as strong projective measurements that collapse the qubits in the measured basis removing their entanglement with the rest of the system that had been generated by the unitary gates (extensions of this to weak/partial measurements are also possible~\cite{Szyniszewski2019,2021arXiv210208381B}). 
While there has been a resurgence of investigation of this question, the idea of quantum to classical phase transitions driven by noise has a long history going back to an early work by Aharonov~\cite{aharonov2000quantum} from over 20 years ago.
Specifically, we consider the model introduced by~\cite{PhysRevB.98.205136,Skinner2019} consisting of alternating circuit layers with random unitary gates and measurement layers in which each qubit is projectively measured with probability $p$, which reduces to random circuit dynamics for $p=0$. In the other extreme limit, $p=1$, the system's state is repeatedly collapsed to an un-entangled product state.

At first glance, one might expect that any non-vanishing measurement probability $0<p\leq 1$ would eventually collapse the system into a short-range entangled after sufficient time evolution. For example, bipartite entanglement $S_A$ between a region $A$ and its complement $A^c$ is generated only by local gates that straddle the boundary $\d A$, and is generated at rate $\sim |\d A|$, whereas the rate of measurement-induced collapse is extensive $\sim p |A|$~\cite{PhysRevB.99.224307}. However, this naive argument ignores a critical feature of the random circuit evolution: scrambling. Namely, random circuit evolution tends to obscure the information stored in a single qubit by encrypting it in a random highly-entangled superposition of many qubits. Consequently, measuring any single qubit in $A$ does not reveal one qubit's worth of information about the state of $A^c$ (as for measuring half of a simple EPR pair), but rather only reveals partial information that roughly scales with the mutual information between the measured qubit and $A^c$: $\mathcal{I}(x) = S_x+S_{A^c}-S_{x\cup A^c}$, where $x$ denotes the distance of the measured qubit from the boundary $\d A$. In the highly entangled states generated by random circuit evolution with local gates, this mutual information generically falls to zero at large distances. 

\begin{figure}[tb!] 
    \centering
    \includegraphics[width=0.85\textwidth]{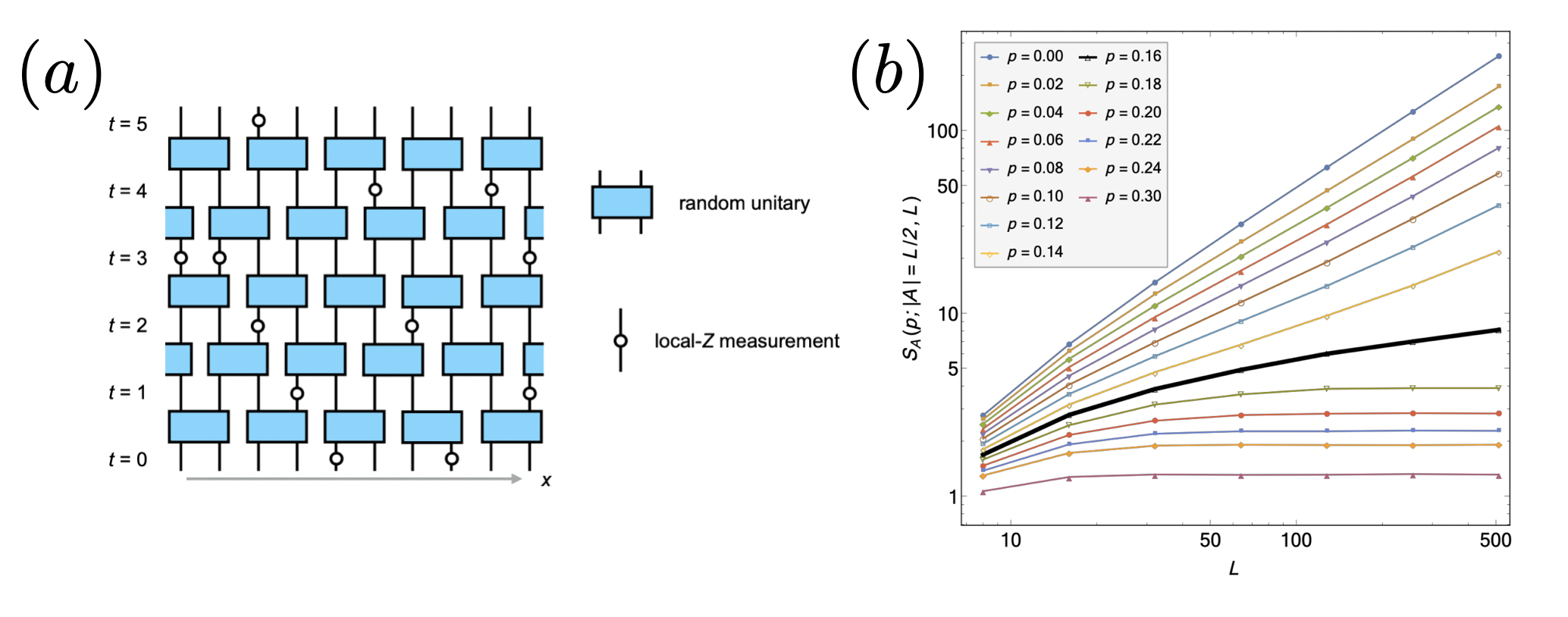}
    \caption{ {\bf Measurement-induced entanglement phase transition } (a) in a monitored random circuit (MRC). (b) Finite-size scaling of measurement-trajectory averaged bipartite entanglement entropy, $S_A$ for region of size $|A|=L/2$, for random Clifford circuits at long times shows a continuous phase transition at critical measurement rate $p_c\approx 0.16$, with volume-law entanglement $S\sim L$ for $p<p_c$, and area-law entanglement $S\sim \text{const.}$ for $p>p_c$. At criticality, long-time entanglement grows logarithmically in $L$ and (and also builds up logarithmically with time). Reproduced from Ref.~\cite{Li2019}.
    \label{fig:mip_model} 
            }
\end{figure}

Considering a $1d$ circuit with $A = (-\infty,0]$ for concreteness, a back-of-the-envelop estimate suggests that if $\mathcal{I}(x)$ falls off faster than $1/x$, then the rate that measurements reduce $S_A$ is $\sim p\int_{-\infty}^0  \mathcal{I}(x) dx$ is a finite constant. Some approximate mean-field-like arguments suggest $\mathcal{I}(x)\sim x^{-3/2}$ in this context~\cite{fan2020self,Li2020b}, and numerically, one finds $\mathcal{I}(x)\sim x^{-1.25}$ compatible with the interpretation of this quantity in terms of the return probability of a directed polymer in a random environment~\cite{2021arXiv210513352L}. Therefore, despite the extensive fraction $\sim p$ of qubits measured after each circuit layer, only measurements close to the boundary of $A$ remove an appreciable amount of entanglement, and the rate of entanglement production by circuit dynamics and entanglement loss due to measurement collapse both scale like $|\d A|$, such that the outcome of their competition depends on $p$. These simple arguments predict the existence of a sharp measurement-induced phase transition (MIPT) at critical measurement probability $0<p_c<1$, with the scrambling dynamics producing highly-entanglement states for $p<p_c$, and measurements collapsing the system into short-range entangled states for $p>p_c$. Indeed, clear numerical evidence for such a measurement-induced entanglement phase transition has been observed in this model for large-scale random Clifford circuits, and smaller-scale circuits with Haar random gates via exact diagonalization~\cite{Li2019}.

\subsection{Entanglement Transition}
The hallmark of the entanglement MIPT in this model is a singular change in the ``average" entanglement entropy, $S(\ell)$, for of a continuous region of size $\ell$ of the typical state produced by the monitored random circuit from volume law, $S(\ell)\sim \ell$ for infrequent measurements ($p<p_c$) to area law ($p>p_c$). Specifically, referring to the output, $|\psi_{\bf m}(t)\>$, of of a particular instance of the random circuit with measurement outcomes ${\bf m}$ as a \emph{trajectory}, we define the trajectory-averaged (Renyi) entanglement entropies for region $A$ as:
\begin{align}
\overline{S^{(n)}_A} = \mathbb{E}_{U,\bf{m}}\[
\frac{1}{1-n}\log\text{tr}_A\(\rho_{A,\mathbf{m}}^n(t)\)
\] ,
\end{align}
where $n$ is the Renyi index (von-Neumann entropy is defined through the limit $n\rightarrow 1$), $\mathbb{E}_\mathbf{m}$ denotes averaging over measurement outcomes (weighted by the Born-probability of obtaining that outcome) and measurement locations, and $\rho_{A,\mathbf{m}}(t) = \text{tr}_{A^c}|\psi_{\mathbf{m}}(t)\>\<\psi_{\mathbf{m}}(t)|$ is the reduced density matrix for the trajectory. 

To be specific, consider the trajectories produced by evolving initially unentangled product states by monitored random circuit evolution with $1+1d$ connectivity, and choose entanglement interval $A$ to be a single contiguous interval of length $\ell$. Numerical simulations~\cite{Li2019} (Fig.~\ref{fig:mip_model}b) show that for $p<p_c$ $\bar{S}(\ell,t)$ grows linearly ($\sim t$) until a time-scale of $t\gtrsim \ell $ where it saturates to a volume law behavior $\bar{S}(\ell) \sim s(p) \ell +\dots $, with volume law coefficient $\log 2 \geq s(p)> 0$ and where $(\dots)$ denotes further subleading-in-$\ell$ terms, including universal $\ell^{1/3}$ corrections~\cite{2021arXiv210513352L}. By contrast, for $p>p_c$, the entanglement quickly saturates to an area law behavior $\bar{S}_A\sim |\d A|$ in $O(1)$ time. Precisely at the transition, $p=p_c$, the entanglement appears to grow logarithmically in time $\bar{S}_A(t)\sim \log t$, saturating to $\bar{S}_A(t\gg |A|)\sim \log \ell$.

The collapse of numerical data for bipartite entropy $\ell = L/2$ over a range of $p$ and system size $L$, are consistent with a universal scaling function:
\begin{align}
\bar{S}(\ell) = G\[(p-p_c)L^{1/\nu},t/L^z\] + \bar{S}_\text{non-universal},
\end{align}
with scaling exponents $z\approx 1$, and $\nu\approx 1.3$ for random Clifford circuits and universal scaling function $G(x,\tau)$ coexists with a non-universal area law background $ \bar{S}_\text{non-universal}$ that evolves smoothly across the transition. For $e^L\gg t\gg L$, the entanglement saturates to a steady-state value with scaling form:
\begin{align}
G(x,\tau\gg1) \sim 
\begin{cases}
	|x|^{\nu} &x\rightarrow -\infty~~\text{(entangling-phase)}
	\\
	\alpha \log(|x|) &x \ll 1~~\text{(critical-regime)}
	\\
	0 & x\rightarrow +\infty~~\text{(collapsed-phase)}
\end{cases}
\end{align}
Critical phenomena aficionados may notice that the critical exponents $\nu,z$ are suspiciously close to those of a $1+1d$ bond-percolation transition $\nu_\text{percolation}=4/3$, $z_\text{percolation}=1$. Indeed the entanglement for monitored random circuits can be mapped to percolation-like statistical mechanical models (see below), however, the transition is believed to be generically different from simple percolation except in the limit of infinite on-site Hilbert space dimension.
For completeness, we note that a distinct but related entanglement transitions between thermalizing and many-body localized (MBL) arises in the purely unitary dynamics (no measurements) generated by a constant or time-periodic Hamiltonians without temporal randomness~\cite{parameswaran2017eigenstate,nandkishore2015many,abanin2019colloquium}. In this context, percolation-type entanglement critical phenomena have also been observed in random Clifford models~\cite{chandran2015semiclassical}.

 A subtle, but crucial point is that the entanglement transition is only visible if one first computes the entanglement of a trajectory, and then averages over trajectories. By contrast, the trajectory-averaged state $\overline{\rho}=\mathbb{E}_{\mathbf{m}}(\rho_{\mathbf{m}})$, is generically volume law entangled for any $p$ (including $p=1$!). As a corollary, the entanglement transition is not visible in averages of local operators: $\mathbb{E}_{\mathbf{m}}\[\<O\>\] = \text{tr}\overline{\rho} O$, but only in their higher moments such as: $\mathbb{E}_{\mathbf{m}}\[\<O_1\>\<O_2\>\]$. This poses a significant challenge to experimentally observing measurement-induced phase transitions: since computing non-linear functions of a trajectory (entanglement, higher-moments of observables etc...) requires measuring many copies of the same trajectory $|\psi_{\mathbf{m}}\>$. Since we cannot simply copy this state (no-cloning theorem!), one must instead sample many-times from the circuit to obtain multiple copies with the same measurement outcomes $\mathbf{m}$. For such non-linear functions of state, this post-selection on measurement outcomes generically adds sampling overhead that is exponential in the space-time volume of the circuit: $\sim \exp(pLt)$ for a circuit of depth $t$ acting on $L$ qubits.

We will discuss below, a possible route to avoiding post-selection through measuring different types of observables involving an ancillary reference qubit initially entangled with the system, designing a classical decoder to avoid the need to prepare multiple copies of a trajectory to detect the area-law phase. Using this strategy, moderate-scale experiments have been successfully performed on trapped-ion experiment~\cite{noel2021} by identifying a model in which the MIPT occurs at very low measurement-density.
We note that these methods are currently specific to Clifford circuits and it is not clear what their overhead would be for MRCs with computationally universal gate sets.
We also note that in recently investigated space-time-duals of RC dynamics~\cite{PhysRevLett.126.060501} (or equivalently in random tensor network states in which the tensors are generate by random circuits~\cite{holographicMPS,2021arXiv210509324C}) a closely related entanglement transition arises which can be observed by postselection only on the \emph{final} measurements of physical qubits in the Bell-basis~\cite{PhysRevLett.126.060501}, incurring exponential overhead only in spatial volume ($\sim e^{L}$ cost) rather than space-time volume ($\sim e^{pLT}$ cost).

\subsection{Alternative perspectives on MIPTs}
Above, we considered a setup where initially un-entangled states were evolved under MRC dynamics resulting in either extensive entanglement production $(p<p_c)$ or continuation of short-range entanglement due to measurement collapse ($p<p_c)$. 
A fruitful alternative perspective if we instead consider feeding mixed states into MRCs (or equivalently states that initially share entanglement with other degrees of freedom), which will expose intriguing connections between the entanglement transition with themes from quantum- information, communication, and error-correction.

\subsubsection{Purification transition}
First consider the trajectories arising from inputing a maximally mixed (``infinite temperature") state $\rho_\infty = \frac{1}{2^L}\mathbbm{1}$ with entanglement $S = -\text{tr}\rho\log \rho=L\log 2$ into an MRC. For $p=1$, every qubit is measured and the state immediately ``purifies" the maximally mixed into a pure quantum state with vanishing total entanglement entropy. By contrast, for $p=0$, the purely unitary RC evolution does not effect $S$, and the state remains maximally mixed at all times. Given our experience with pure-state inputs, it is natural to expect that there is a critical measurement probability $p_c$, where MRC dynamics undergoes a phase transition between regimes where densely-repeated measurements purify any mixed initial state ($p>p_c$) or fail to do so due to scrambling dynamics which obscure whether the measured qubit is in a mixed state due to environmental entanglement or entanglement other qubits in the system ($p<p_c$). A priori it might not be  obvious that the purification transition~\cite{Gullans2019} for mixed state inputs should coincide precisely with entanglement transitions, however numerically they appear to do so~\cite{Gullans2019}, and we will see below in the context of statistical mechanical mappings that these transitions have a unified ``dual" interpretation as the same bulk-ordering transition of a replica-spin model. 

An important caveat to the purification interpretation is that at ultra-long time scales ($t\sim \exp L$), MRCs for any non-zero measurement fraction ($p>0$) will eventually purify an arbitrary input state.  Hence, the purification transition is only evident in the limit $e^L \gg t \gg L \gg 1$.

\begin{figure}[tb!] 
    \centering
    \includegraphics[width=0.85\textwidth]{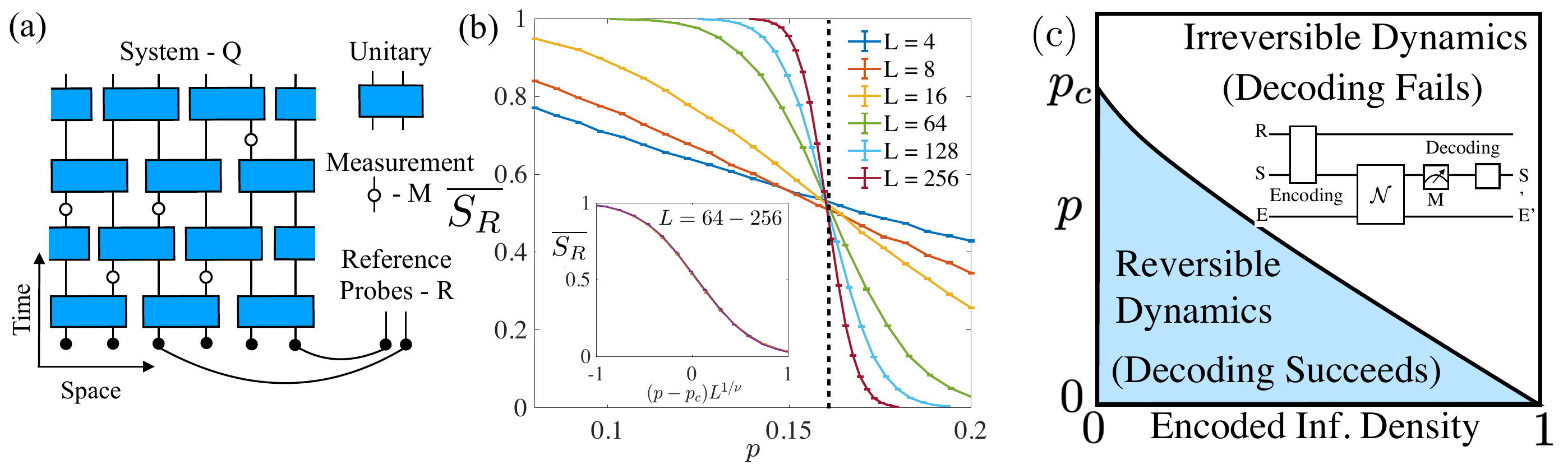}
    \caption{ {\bf Quantum information perspectives on the MIPT entanglement transition -- } adapted from \cite{Gullans2019}. (a) Entanglement transition as a purification transition for entangled reference ancilla qubits (R). The entanglement entropy of the reference, $\overline{S_R}$, serves as an order parameter for the transition. When a pre-scrambling unitary is included before the MRC dynamics, $\overline{S_R}$ jumps from $\log(2)$ to $0$ in the infinite size limit as evidenced by finite size scaling shown in (b) for random Clifford circuits. (c) Shows an alternative quantum communications perspective of this setup, where the pre-scrambling circuit is viewed as a random encoding of the quantum information between R and the system (S), the MRC channel $\mathcal{N}$ represents a communications channel, including monitoring by an environment $E$, and then one attempts to decode the information at the output (with a hypothetical optimal decoder). The entanglement transition represents a phase transition in the quantum channel capacity of this communications setup (c).
    \label{fig:ancilla} 
            }
\end{figure}

\subsubsection{Ancilla probe of purification transition}
We can view the maximally mixed input state $\rho_\infty$ as arising from having each qubit in the system being entangled with an ancilla qubit that does not participate in the circuit dynamics. The purification perspective then suggests a useful way to characterize the entanglement/purification transition via examining whether the mutual information between ancillas and the system (S) survives to long times, or is killed by measurement-collapse. 

In fact, to observe the transition, it suffices to examine just a single ``reference" ancilla, $R$, and look at the trajectory-average of the reference ancilla~\cite{Gullans2020}:
\begin{align}
\overline{S_R} = \mathbb{E}_{U,m}\[S_R\] \equiv \mathbb{E}_{U,m}\[ -\text{tr}_R~\rho_R\log\rho_R\].
\end{align}
Measured at times $2^L\gg t \gg L$, and in the limit $L\rightarrow \infty$, $\overline{S_R}$ exhibits a discontinuous jump across the transition. This jump provides a convenient numerical signature that precisely locates the transition via a crossing  in curves of $\overline{S_R}$ versus $p$ for different $L$~\footnote{As an aside, we note that, if the ancilla qubit, $R$, is initially entangled non-locally with the system, e.g. by applying a scrambling unitary before undergoing MRC dynamics, then in the $L \rightarrow \infty$ limit, $\overline{S_R}$ precisely jumps from $\log 2$ for $p<p_c$ to $0$ for $p>p_c$. On the other hand, if the ancilla qubit is locally entangled with a single system qubit, $\overline{S_R}$ is not quantized (for example with probability $p$ that qubit could immediately get measured even for $p<p_c$) and its jump across the transition is non-universal.}.
This single-ancilla feature has been referred to as a \emph{scalable} probe of MIPT, in the sense that it avoids the exponential-in-$L$ cost of measuring many-body entanglement of a trajectory.

\subsubsection{Experimental observation of MIPT in trapped-ions}

Using a standard methods of measuring the reference qubit entanglement entropy (e.g. using tomography to determine its density matrix) would still require many copies of a given trajectory which would incur a much larger post-selection overhead.
However, Refs.~\cite{Gullans2019,noel2021} highlight a method to potentially avoid post-selection. In the purifying phase, the measurements collapse the reference ancilla into a pure state disentangled from the system. This pure state may be in a random basis determined by various measurement outcomes, so that further measurements of the ancilla in a fixed (e.g. computational) basis would generally see a large-entropy mixture of $0$ and $1$ outcomes. However, if this basis can be determined using the knowledge extracted from the measurement outcomes in the circuit, then one could observe the purity of the entangled state without preparing multiple copies of the same trajectory. This idea can be carried out for Clifford circuits~\cite{noel2021}, whose classically efficient simulations permit one to design a feedforward circuit using quantum logic to transform the ancillas into the computational basis when they are purified. Using this technique, Ref.~\cite{noel2021} was able to observe finite-size signatures of a MIPT in a trapped-ion chain  without post-selection. While this experiment was relatively small scale, involving 8 system qubits and one reference qubit and a variable small number ($\leq 4$) of measurements, clear signs of the limiting behavior in the large- and small- measurement regime were observed, and the methods developed pave the way for larger scale experiments (e.g. in architectures where mid-circuit measurements can be performed to avoid the need to sacrifice ancilla qubits as a classical register to hold measurement outcomes). 
Recent work~\cite{yoshida2021decoding} further shows that related ideas can allow detection of the scrambling phase without postselection through entanglement distillation methods.

\begin{figure}[tb!] 
    \centering
    \includegraphics[width=1.0\textwidth]{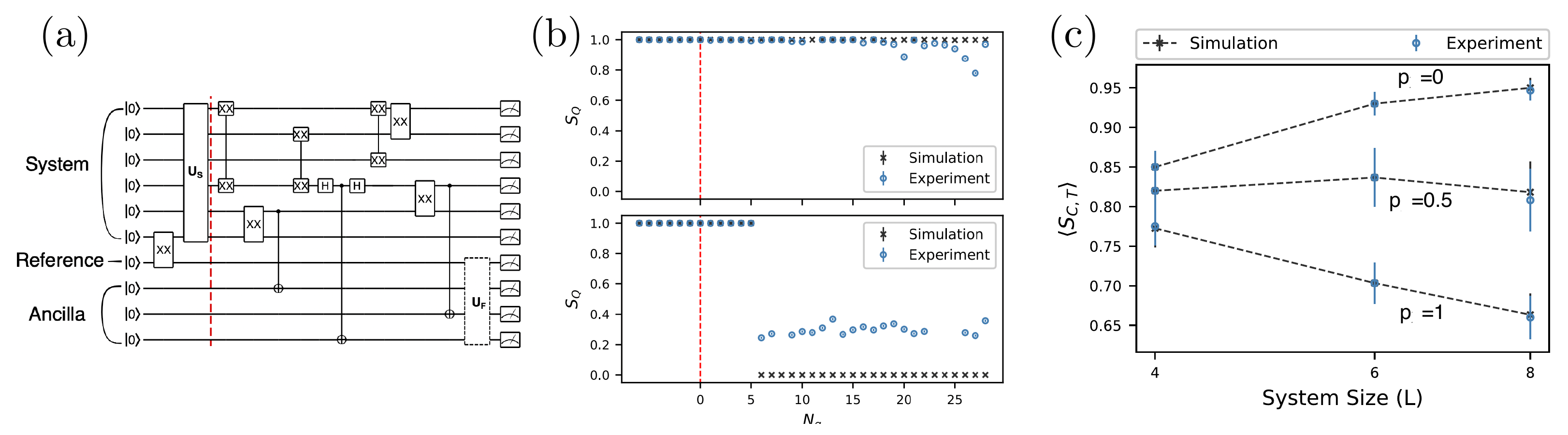}
    \caption{ {\bf Finite-size evidence of MIPT entanglement transition in trapped ion chain -- } adapted from \cite{noel2021}. (a) Circuit schematic of experiment: a reference qubit R is entangled with the system S through an entangling XX-gate (unitary $u_{XX} = e^{i\pi/4 X\otimes X}$). The information is then scrambled with scrambling circuit $U_C$, and subjected to random Clifford MRC dynamics. Four ancilla A hold the intermediate measurement outcomes, deferring measurement to the end of the circuit. Before measurement, a feedback decoding circuit $U_F$ is then applied to the R and A to attempt to disentangle the reference and measurement-ancilla and place the reference in the computational basis (which will be successful if the ancilla is purified). (b) Quantum entanglement $S_Q$ of R determined through tomography and averaged over $\sim 10^{3-4}$ shots for two trajectories where R stays mixed (top) or is purified (bottom). (c) Classical entropy of R after a thresholding procedure, $\<S_{C,T}\>$, for various system sizes and measurement rates show a qualitative change in system-size dependence across the MIPT (determined by simulation to occur at $p\approx 0.72$).
    \label{fig:expt} 
            }
\end{figure}

We emphasize that the efficient measurement-decoding implementations in these works are special to Clifford circuits and exploited their efficient classical simulability. While Clifford circuits are in interesting an important class of quantum circuits that play an important role in many areas of quantum information such as stabilizer codes (among others), they are also in a sense fine-tuned and non-generic in that they are not capable of universal quantum computation, and that small deviations from Cliffordness ultimately spoil their efficient simulation. It remains an important open question whether similar decoding strategies can be used to reduce the overhead associated with post-selection to observe MIPTs in MRCs with universal gate sets.

\subsubsection{Connection to quantum channel capacity and quantum error correction}
The purification perspective also suggests intriguing connections between MIPTs to quantum communication and error correction~\cite{Choi2020,Gullans2019}.
Specifically, one can view the MRC as a communication channel transmitting input mixed state $\rho_\text{in}$ initially entangled with reference system $R$ through the MRC dynamics to an output. 
A key metric for the quality of a quantum communication channel is quantum channel capacity (QCC), which is roughly-speaking the number of qubits'-worth of information that can be transmitted through a noisy channel, optimized over all possible encoding and decoding protocols. 
A famous result~\cite{smith2010quantum} of Lloyd, Shor, and Devetak [LSD] relates the QCC for a channel $\mathcal{N}$ to the coherent information: $I_C[\mathcal{N},\rho_\text{in}] = S_{A'} - S_{A'R}$ where $S_X$ is the von-Neumann entanglement entropy of subsystem $X$, $\rho_\text{in}$ is the input state of system $A$, $A'$ denotes the system qubits after transmission through the channel, and $R$ is the reference ancilla system with which $A$ is initially entangled, and which does not undergo any dynamics. This quantity can be understood as the amount of entanglement that survives from input to output by subtracting off ``incoherent" entanglement with the environment responsible for information loss in the channel. To see this note that we can purify the channel as a unitary interaction between the system and an environment $E$, and  $S_{A'R} = S_E$ where $E$ is the environment that from the output. Namely LSD showed that the quantum channel capacity is equal to the stabilizing limit of the coherent information maximized over input state:
\begin{align}
\text{QCC} = \lim_{n\rightarrow \infty} \frac{1}{n}\max_{\rho_\in} I_C\[\mathcal{N}^\otimes{n},\rho_\text{in}\].
\end{align}
Intuitively, taking a large number of copies reflects that channel capacity characterizes the ability to transmit long sequences of communications rather than a single message. For a certain class of so-called ``degradable" quantum channels, which generalize dephasing and erasure error channels, and which were shown~\cite{Choi2020} to include MRCs, the coherent information is simply additive, $I_c(\mathcal{N}^{\otimes n}) = nI_c(\mathcal{N})$, and it suffices to consider just a single copy of the channel. Further, the average coherent information for MRCs was shown~\cite{Choi2020} to be precisely equal to $\overline{S_R}$.

This strongly suggests that the entanglement/purification transition in MIPTs can also be regarded as a phase transition in quantum channel capacity (QCC) between a high-capacity phase where the MRC-channel transmitting an extensive number of qubits ($p<p_c$) and a low-capacity phase where with vanishing fraction of qubit-information transmitted ($p>p_c$). 

Strictly speaking,  $\overline{S_R}$ is not precisely the same as QCC, but rather reflects a sort of ``average" QCC. Namely, the typical formal definition of QCC involves optimizing over input state $\rho_\text{in}$ for each random circuit instance (the encoding of the input may exploit specific knowledge of the circuit-gates and measurement locations, e.g. to avoid encoding information near positions that are heavily measured), and average QCC would be obtained by averaging each optimized result over circuit configuration. Instead, $\overline{S_R}$ captures a different order of limits, where averaging over the MRC-ensemble is performed before optimization over inputs (it can be shown that the optimal input to the average channel is the maximally-mixed state $\rho_\infty$~\cite{Choi2020}). 
We note however, that optimal communication capacity is rarely if ever achieved in practice, and a more physically relevant question is whether there exists a threshold error-rate below which one can communicate encoded information at a finite rate.
The above results (see also further detailed arguments and numerical simulations in~\cite{Choi2020,Gullans2019}) show that MRCs indeed have such a threshold.

The measurement-induced entanglement transition can also be understood as a quantum error-correction (QEC) threshold~\cite{Choi2020,PhysRevX.11.031066}. Indeed, the QCC problem, which involves encoding information into a quantum state, and subjecting it to a noisy and error-prone channel before attempting to recover it using a decoder is a form of QEC.

\subsubsection{Information gained by the observer}

A closely related perspective of the MIPT can be formulated in terms of the  ability of the observer to extract information about the state of the system. In the volume-law phase, information scrambling by the unitary evolution hides information into highly non-local degrees of freedom that are hidden from the local measurements. As a result, in that phase, the observer would require a time scaling exponentially with the system size in order to extract all the information about the system. In contrast, in the area-law phase, the observer can learn everything about the state of the system in a time of order one. The amount of information extracted by the measurements about the initial state of the system can be quantified by the Fisher information~\cite{Bao2020}, which is non-analytic at the MIPT.

\section{Replica statistical mechanics models} \label{secStatMech}

Most of the phenomenology of measurement-induced phase transitions described above relied on numerical results, either on small ($L \sim 20$ qubits) Haar circuits or larger ($\sim 10^3$ qubits or qudits) Clifford circuits. To understand the scaling properties and phase structure  of monitored quantum circuits on a firmer ground, we now turn to an analytic approach by deriving an exact mapping onto a statistical mechanics model. Using a replica trick, entanglement properties can be mapped onto the free energy cost of a boundary domain wall in a classical ``spin'' model~\cite{PhysRevB.100.134203,Zhou2019, Bao2020,Jian2020}, with the entanglement transition corresponding to a simple (replica) symmetry breaking transition. 
This replica approach to performing statistical mechanics mappings for entanglement transitions was first introduced in the context of random tensor network states~\cite{PhysRevB.100.134203}, which turn out to be very closely related to MIPTs in MRCs.

\subsection{Replica trick}

Our goal is to compute the Renyi entropies of such individual quantum trajectories, averaged over measurement outcomes and random unitary gates. Each trajectory is weighted by the Born probability $p_\mathbf{m}={\rm tr} \rho_{\mathbf{m}}$ where $\rho_{\mathbf{m}} = \left| \psi_{\mathbf{m}}  \rangle \langle  \psi_{\mathbf{m}} \right|$ is the (pure) density matrix of the system:
\begin{equation} \label{eqDefRenyi}
S_{A}^{(n)} =\mathbb{E}_{U} \sum_{\mathbf{m}} p_\mathbf{m} \frac{1}{1-n}  \log\[ \frac{{\rm tr} \rho^n_{A,{\mathbf{m}}}}{({\rm tr} \rho_{\mathbf{m}})^n}\],
\end{equation}
where $\mathbb{E}_{U} $ refers to the Haar average over random unitary gates, and $\sum_{\mathbf{m}}$ denotes averaging over quantum trajectories ({\it i.e.} over measurement outcomes). 
Here, it will turn out to be convenient to work without an explicitly normalized density matrix in computing time evolution, and so we include the factors of $\sim \text{tr}\rho$ in the denominator to explicitly enforce normalization of the density matrix when computing observables.
We will denote $\overline{S_{A}^{(n)}}$ the Renyi entropy averaged over measurement locations. 

On the face of it, computing the Renyi entropies~\eqref{eqDefRenyi} might seem like a daunting task: entanglement properties are usually hard to access analytically, and the non-equilibrium time evolution combined with the non-linearity of the measurements make the problem even harder. However, following Refs.~\onlinecite{PhysRevB.100.134203,Zhou2019, Bao2020,Jian2020}, we can use the replica trick to compute~\eqref{eqDefRenyi}. As in the field of classical random spin models and spin glasses, the basic idea is to rely on the simple identity:
\begin{equation} \label{eqDefRenyi}
\log x = \underset{k \to 0}{\rm lim} \frac{x^k-1}{k}.
\end{equation}
This equation is exact, but the ``trick'' is to compute the average of the logarithm $\log x$ (here over random unitaries and measurement outcomes), which is a hard task in general, using the average of the moment $x^k$ where $k$ is an integer, which is usually a lot easier. This step involves analytic continuation in $k$, which can be subtle in some cases. 
Using this replica trick, we can write the Renyi entropies as
\begin{equation} \label{eqSswap}
S_{A}^{(n)} =\underset{k \to 0}{\rm lim} \ \mathbb{E}_{U} \sum_{\mathbf{m}} \frac{p_{\mathbf{m}} }{(1-n)k}  \left( ({\rm tr} \rho^n_{A,{\mathbf{m}}})^k - ({\rm tr} \rho_{\mathbf{m}}^{\otimes k n })  \right).
\end{equation}
We will write $Q=nk  +1$ to denote the total number of replicas, where the additional replica comes from the Born probability $p_{\mathbf{m}}={\rm tr} \rho_{\mathbf{m}}$ weighting different quantum trajectories. Within this replicated state, we can write
\begin{equation} 
S_{A}^{(n)} =\underset{k \to 0}{\rm lim} \ \frac{1 }{(1-n)k} \mathbb{E}_{U} \sum_{\mathbf{m}} \left( {\rm tr} \left[{\cal S}^{\otimes k}_{A,n} \rho_{{\mathbf{m}}}^{\otimes Q}\right]- {\rm tr} \left[\rho_{\mathbf{m}}^{\otimes Q } \right]  \right),
\end{equation}
where ${\cal S}_{A,n}$ is a permutation ``swap'' operator implementing the partial trace in the region $A$, acting on each of the first $k$  replicas (which are themselves  $n$-fold replicated states) as
\begin{equation}\label{eq:boundary def}
{\cal S}_{A,n}=\prod_{x} \chi_{g_x},\quad
g_x=\left\{\begin{array}{ll}(12\cdots n), & x\in A,\\
{\rm identity} = e,
& x\in \bar{A}.\end{array}\right.
\end{equation}
$g_x$ labels the permutation on site $x$, and ${\chi}_{g_x}=\sum_{[i]}\ket{i_{g_x(1)}i_{g_x(2)}\cdots i_{g_x(n)}}\bra{i_1i_2\cdots i_n}$ is its representation on the replicated on-site Hilbert space,
i.e. on its $n$-fold tensor product. As indicated in the equation above, $g_x$ is the cyclic (identity) permutation when $x$ is in the region $A$ (when $x$ is in the region $\bar{A}$).  Here, we use standard cycle notation to denote permutations, for example $(123)4$ refers to the cyclic permutation $1234\rightarrow 2314$.

\subsection{Haar calculus and Boltzmann weights}

\begin{figure}[bt!] 
    \centering
    \includegraphics[width=0.7\textwidth]{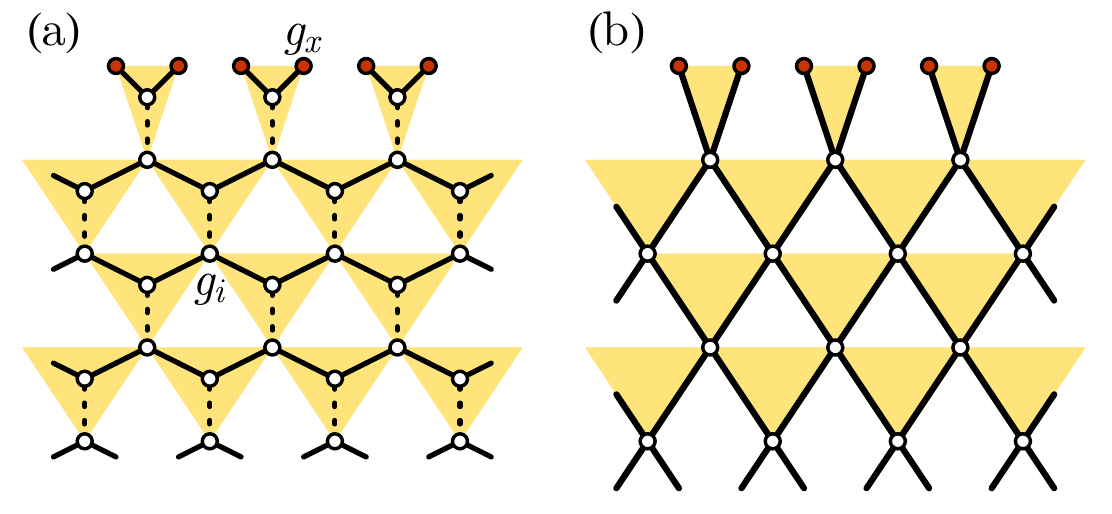}
    \caption{ {\bf Statistical mechanics model } (a) Geometry of the statistical mechanics model of $S_Q$ spins. The red sites corresponds to the boundary spins to be pinned by the boundary condition. (b) In the $d=\infty$ limit, the model reduces to a Potts model on a square lattice. Reproduced from Ref.~\cite{Jian2020}.
    \label{fig:statmech} 
            }
\end{figure}

The next step is to perform the average over the (replicated) random unitary gates, using the Haar measure (see~\cite{roberts2017chaos} for a nice physicist-accessible review of some technical aspects of Haar averages over the unitary group). The average over each gate can be evaluated using the formula
\begin{equation} \label{eqHaaraverage}
\mathbb{E}_{U\in U(D)}\left(U^{*Q}\otimes U^{Q}\right) = \sum_{\sigma,\tau \in S_{Q}} \text{Wg}_{D}(g_1^{-1}g_2){\mathcal X}_v (g_1) \otimes{\mathcal X}_v (g_2),
\end{equation}
where $g_1,g_2$ are permutations of the replicas, Wg are called Weingarten functions, $D=d^2$,  ${\mathcal X}_v (g_1)={\mathcal X} (g_1) \otimes{\mathcal X} (g_1)$  permutes the output legs of $U$ by $g_1$, and contracts them with the corresponding legs of $U^*$, and similarly for ${\mathcal X}_v(g_2)$ acting on the input legs. The reason the average $\mathbb{E}_{U\in U(D)}\left(U^{*Q}\otimes U^{Q}\right) $ can be expanded onto permutations of the replicas is that such permutations commute with the action of the unitaries, which are the only terms surviving the Haar average. The commuting actions of the unitary and permutation groups on a tensor product Hilbert space is a mathematical statement known as {\em Schur-Weyl duality}. This step can be generalized to other subgroups of the unitary groups, see Ref.~\cite{2021arXiv211002988L}. Using standard tensor network notations~\cite{ORUS2014117}, we will write eq.~\eqref{eqHaaraverage} as~\cite{Jian2020}
\begin{equation} \label{eqHaaraverage2}
\avg_U \dia{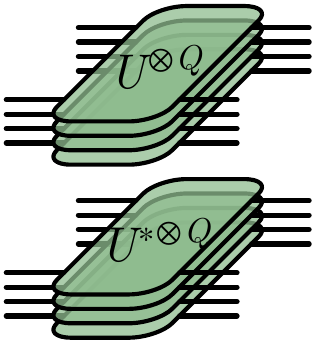}{56}{-26}=\sum_{g_1,g_2\in S_Q}\text{Wg}_{D}(g_1^{-1}g_2)\dia{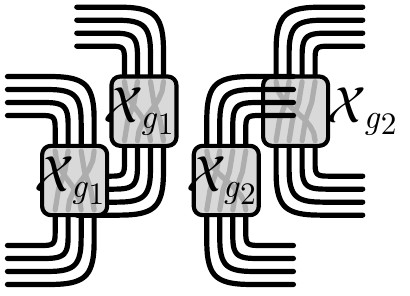}{47}{-20},
\end{equation}
Here $\text{Wg}_{D}(g)$ denotes the Weingarten function of the permutation $g$, 
\begin{equation}\label{eq:Wg def}
\text{Wg}_D(g)=\frac{1}{Q!}\sum_{\lambda\vdash Q}\frac{\chi_\lambda(e)\chi_\lambda(g)}{\prod_{(i,j)\in Y(\lambda)}(D-i+j)},
\end{equation}
where the sum is taken over all integer partitions $\lambda$ of $Q$ [denoted in the above equation by the notation $\lambda\vdash Q$, such that $\lambda = (\lambda_1, \lambda_2, ...)$ with $\lambda_1\geq \lambda_2\geq\cdots$, $\lambda_i\in\mathbb{N}$ and $\sum_i\lambda_i=Q$],
and the product is taken over all cells $(i,j)$ in the Young diagram $Y(\lambda)$ of the shape $\lambda$. Here $e$ denotes the identity group element, and $\chi_{\lambda}(g)$ is the irreducible character of the symmetric group $S_Q$ indexed by the partition $\lambda$. Those Weingarten coefficients $ \text{Wg}_{D}(g)$ can also be computed by contracting the unitaries within each replica in the left-hand side of eq.~\eqref{eqHaaraverage} to obtain a trivial result using $U^\dagger U =1$. This yields
\begin{equation}
{\mathcal X}_v (e) = \sum_{g_1,g_2 \in S_{Q}} \text{Wg}_{D}(g_1^{-1}g_2) \left( {\rm tr}{\mathcal X}_v (g_1)\right){\mathcal X}_v (g_2),
\end{equation}
where $e$ is the identity permutation. The trace of ${\mathcal X} (g)$ simply counts the number of {\em cycles} in the permutation $g$, 
\begin{equation}
 {\rm tr}{\mathcal X}_v (g) = D^{C(g)},
\end{equation}
where $C(g)$ is the number of cycles in the permutation $g$. We thus find
\begin{equation}
\sum_{g_1 \in S_{Q}} \text{Wg}_{D}(g_1^{-1} g_2) D^{C(g_2)} = \delta_{g_2}, 
\end{equation}
where $\delta_g$ is equal to $1$ is $g=e$, and 0 otherwise. This equation can be used to define the Weingarten coefficients $\text{Wg}_{D}$, as the inverse of  $D^{C}$. Equation~\eqref{eqHaaraverage} apply to the brick-wall pattern of unitary gates defines a statistical model on the honeycomb lattice, where permutations live on vertices. Contracting unitary gates can be done by assigning a weight to links connecting unitaries given by 
\begin{equation} 
W(g_1, g_2) = {\rm tr} \big [{\mathcal X}^\dagger(g_1){\mathcal X}(g_2)\big ]   = d^{C(g_1^{-1}g_2)}.
\end{equation}
Note the factor of $d$ here, instead of $D$, since we are focusing on a single leg of the unitary (${\mathcal X}_v = {\mathcal X} \otimes {\mathcal X}$). This weight is associated to all links that were not measured. If a link was measured instead, all replicas are constrained to be in the same state, and the weight is instead $d$ after averaging over possible measurement outcomes. Those equations fully determine the weights of the statistical model in monitored Haar random circuits. Upon averaging over measurement locations and outcomes, the weight assigned to a link is therefore given by~\cite{Jian2020}
\begin{equation}
W_p(g_1, g_2) = (1-p) d^{C(g_1^{-1}g_2)} + p d.
\end{equation}
Putting these results together and ignoring for the moment boundary conditions, we obtain an anisotropic statistical mechanics model defined on the honeycomb lattice, 
\begin{equation} \label{eqZHaar}
Z = \sum_{\lbrace g_i \in S_Q \rbrace} \prod_{ \langle ij \rangle \in G_{s}}W_p(g_{i}^{-1} g_j) \prod_{ \langle ij \rangle \in G_{d}} \text{Wg}_{D}(g_{i}^{-1} g_j), 
\end{equation}
where $G_s$ ($G_d$) denotes the set of solid (dashed) links on the lattice. In Fig.~\ref{fig:statmech}, the vertical (dashed) links on the honeycomb lattice represent the Weingarten functions which originated from averaging the two-site unitary gates, and the solid links keep track of the link weights originating from averaging over measurements.

\subsection{Boundary conditions and domain wall free energy}

By imposing different boundary conditions corresponding to fixing permutations at the boundary, the statistical mechanics model results in different partition functions
\begin{equation}\label{eq:Z def}
\begin{split}
Z_A&=\mathbb{E}_{U,\bf{m}}  {\rm tr} \left[{\cal S}^{\otimes k}_{A,n} \rho_{{\mathbf{m}}}^{\otimes Q}\right] ,\\
Z_0&=\mathbb{E}_{U,\bf{m}}{\rm tr} \left[\rho_{\mathbf{m}}^{\otimes Q } \right] ,
\end{split}
\end{equation}
from which the averaged $n$th R\'enyi entropy $\overline{S^{(n)}_A} $ can be obtained in the replica limit via
\begin{equation}
\overline{S^{(n)}_A}  =\frac{n}{1-n}\lim_{Q\to 1}\frac{Z_A-Z_0}{Q-1}.
\end{equation}
Using the fact that $Z_A= Z_0 = 1$ in the replica limit $k \to 0$ ($Q \to 1$), this can be rewritten in a more intuitive form as the free energy cost of the domain-wall associated with changing the boundary condition in the entanglement region:
\begin{equation}
\label{eqFenergycost}
\overline{S^{(n)}_A}  =\lim_{k \to 0}\frac{F_{A}-F_0}{k(n-1)},
\end{equation}
with $F_A = - \log Z_{A}$ and $F_0 = - \log Z_{0}$. The $S_Q$ `spins' on the boundary, which are permutation group elements $g_x\in S_Q$ for boundary sites $x$, are pinned by the boundary condition which is uniform and set to $g_x = e$ for $Z_0$, corresponding to a trivial contraction. The partial trace in $Z_A$ is implemented as follows:
\begin{equation}\label{eq:SQ boundary}
g_x=\left\{\begin{array}{ll}g_{\rm SWAP} \equiv (12\cdots n)^{\otimes k}, & x\in A,\\ 
{\rm identity} = e,
& x\in\bar{A}.\end{array}\right.
\end{equation}
This equation follows immediately from $k$ copies of \eqref{eq:boundary def}. 

Now that we have mapped the calculation of the entanglement entropies of the random circuit with projective measurements onto a (replica) statistical mechanics model, many qualitative features of the entanglement transition can be understood naturally.  At small $p$, the Boltzmann weights give a ferromagnetic interaction favoring  group elements on neighboring sites to be equal,
and we thus expect an ordered phase of the statistical mechanics model. In that phase, the free energy cost  $F_A-F_0$ in \eqref{eqFenergycost} associated with changing the boundary conditions in the region $A$, and thus of creating a domain wall, scales with the size of the interval $L_A$ of $A$ at long times, corresponding to volume-law entanglement $ \overline{S^{(n)}_A} \sim L_A$. As the measurement rate $p$ gets closer to 1, the effective temperature of the statistical mechanics model is increased, leading to a disordered phase. The domain wall condensate present in this phase can freely absorb the domain wall at the boundaries of the entanglement interval, such that, for a distance exceeding the correlation length from the boundary, there is no additional free energy cost from the boundary domain. In this limit, the free-energy cost of the boundary domain will scale like the boundary of $A$, corresponding to area-law scaling of entanglement $\overline{S^{(n)}_A}\sim \text{const}$.

\subsection{Symmetry and conformal invariance}

A crucial property of the statistical mechanics model derived above  (eq.~\eqref{eqZHaar}) is that the Boltzmann weights are invariant under global right- and left-multiplication of all group elements
\begin{eqnarray}
\label{LabelEqSymmetry}
g_i \to h^{\vphantom\dagger}_L g_i h_R^{-1},  g_j \to h^{\vphantom\dagger}_L g_j h_R^{-1},
\   {\rm where} \ h_L, h_R \in S_Q.
\end{eqnarray}
This follows from the fact that both the cycle counting function and the Weingarten functions (which are inverse of each other) are {\em class functions}, that is, they depend only the conjugacy class of the permutation group element. Physically, the two factors of $S_Q$ symmetry correspond to the separate invariance under permuting the replicas in the forward- ($U$) and backward- ($U^*$) time contours. This structure will be important in the next section, and in the discussion of MRCs with symmetry below.

This general mapping indicates that the measurement-induced transition corresponds to a simple ordering, (replica) symmetry breaking transition.  In general, assuming that the transition is of second order, it should be described a by two-dimensional Conformal Field Theory (CFT) with central charge $c=0$ in the replica limit $Q \to 1$. (Recall that $c$ measures the way the free energy changes when a finite scale is introduced; since here the partition function $Z_{0}=1$ is trivial in the replica limit, we have $c=0$.) 
Such CFTs at central charge $c=0$ are non-unitary, and are especially complicated even in two dimensions. However, there are a number of general properties that follow from general scaling considerations and conformal invariance. 

Since the bulk properties of the transition only depend on $Q$, the statistical model approach naturally explains why all R\'enyi entropies with $n \geq 1$ have a transition at the same value of $p_c$, as observed in the numerics. Conformal invariance also allows us to derive the general scaling form of the entanglement entropy near criticality, by noting  that the ratio of partition functions  $Z_{A}/Z_{0}$ that appears in~\eqref{eqFenergycost}, corresponds in the CFT language to the two-point function of a boundary condition changing (BCC) 
operator $\phi_{\rm BCC}$~\cite{cardy_conformal_1984,cardy_boundary_2006}
\begin{eqnarray}
\label{LabelEqDefBccOperator}
&&
Z_{A}/Z_{0} = \langle \phi_{\rm BCC}(L_A) \phi_{\rm BCC}(0) \rangle,
\end{eqnarray}
where the operators are inserted at  the boundary of the entanglement interval $A$. Near criticality, this two-point function scales as $\sim 1/L_A^{2 h (n,m)} f_{n,m} (L_A/\xi_{Q})$ with $\xi_Q \sim \left| p - p_c(Q) \right|^{-\nu(Q)}$ the correlation length of the statistical mechanics model and  $f_{n,m}$ are universal scaling functions that depend on $n$ and $m$ independently. Plugging this expression into the replica formula~\eqref{eqFenergycost}, we find the general scaling of the entanglement entropy (up to non-universal additive terms)
\begin{equation}
\overline{S^{(n)}_A}   =  \alpha_n \log L_A + f_n \left( \frac{L_A}{\xi}\right),
\label{eqEntanglementLog}
\end{equation}
with $\xi \sim \left| p - p_c\right|^{-\nu}$ the correlation length in the limit $Q\to 1$, and $\alpha_n = \frac{2}{n-1} \left.\frac{\partial h}{\partial m} \right|_{m=0}$ is a universal prefactor. Note that $\alpha_n$ is unrelated to the central charge of the theory, and instead is a boundary critical exponent. In particular, conformal invariance predicts that $\overline{S^{(n)}_A}    \sim \log L_A$ at criticality $p=p_c$, with a universal prefactor that depends on the R\'enyi index $n$.

\subsection{Large Hilbert space dimension limit}

\subsubsection{Mapping onto classical percolation}

In the limit of large on-site Hilbert space dimension, $d\rightarrow\infty$, the $S_Q$ model above reduces
to a Potts model with $Q!$ colors defined on the square lattice. To see this, we evaluate the partition function weight $J_p(g_i,g_j;g_k)$ associated with
each down triangle in Fig.~\ref{fig:statmech}, integrating out the middle spin:
\begin{equation}\label{eq:J def}
J_p(g_i,g_j;g_k)=\sum_{g_l\in S_Q}W_p(g_i^{-1}g_l)W_p(g_j^{-1}g_l)\text{Wg}_{D}(g_l^{-1}g_k).
\end{equation}
The partition function can then be equivalently written in terms of the triangle weight $J_p$ as
\begin{equation}\label{eq:Zinfd}
Z=\sum_{\{g_i\in S_Q\}}\prod_{\langle ijk\rangle}J_p(g_i,g_j;g_k),
\end{equation}
subject to the appropriate boundary conditions that distinguish $Z_0$ from $Z_A$.
 In the $d\to\infty$ limit, we have $d^{C(g)} \sim d^{Q}\delta_g$, where $\delta_g$ is the delta function that gives 1 if and only if $g=e$ is the identity element in the permutation group $S_Q$, and gives 0 otherwise. This follows from the fact that the number of cycles $C(g)$ is maximized by the trivial permutation: $C(g) = Q$. Since the Weingarten weights are defined as the inverse of $D^{C(g)}$ with $D=d^2$, we immediately find that in the $d\to\infty$ limit, we have $\mathsf{Wg}_{D}(g)=D^{-Q}\delta_g$. Substituting into the triangle weight, the triangle weight~\eqref{eq:J def} and after some straightforward algebra, one finds~\cite{Jian2020}
\begin{equation}\label{eq:J inf d}
J_p(g_i,g_j;g_k)=((1-p)\delta_{g_i^{-1}g_k}+p)((1-p)\delta_{g_j^{-1}g_k}+p),
\end{equation}
which further factorizes into partition function weights defined separately on the bonds $\langle ik\rangle$ and $\langle jk\rangle$. The partition function weight across the bond $\langle ik\rangle$
equals
$1$ if $g_i=g_k$ and $p$ if $g_i\neq g_k$, and  an analogous weight is assigned to the bond $\langle jk\rangle$. If we treat each on-site group element $g_i\in S_Q$ as a state (color) in a spin model, the partition function weight precisely matches that of a $Q!$-state Potts model on a square lattice, whose links are between sites $i$ and $k$, and between sites $i$ and $j$ in each unit cell. 

In order to analytically continue $Q \to 1$, we rewrite the partition function of this Potts model in terms of the so-called Fortuin-Kasteleyn (FK) cluster expansion~\cite{FORTUIN1972536}. We expand the partition function as a product over links with weight $(1-p)\delta_{g_i^{-1}g_k}+p$ , by assigning an ``occupied'' link to the term $(1-p)\delta_{g_i^{-1}g_k}$, while an empty link corresponds to picking the trivial term $p$ in the product. The occupied links form clusters, where the permutation spins are forced to be the same due to the Kronecker delta functions. One can then perform the sum over permutations $\sum_{\{g_i\in S_Q\}}$ in the partition function~\eqref{eq:Zinfd} exactly, which allows us to rewrite it as a sum over FK clusters:
\begin{equation}\label{eq:Zpercolation}
Z=\sum_{\rm clusters} p^{\# {\rm empty \ links}} (1-p)^{\# {\rm occupied \ links}} \left(Q! \right)^{\# \rm clusters}.
\end{equation}
Using this exact rewriting of the partition function, we can now readily take the replica limit $Q \to 1$ since $Q$ only appears in the Boltzmann weight of the clusters in that formulation. In the replica limit, all clusters carry a trivial weight, and the partition function~\eqref{eq:Zpercolation} describes a classical bond percolation problem, where links are occupied with probability $1-p$ (no measurement), and are empty with probability $p$ (corresponding intuitively to a local measurement cutting the circuit). This percolation picture of the transition is rather appealing and natural, and predicts critical exponents that are close to the ones measured even for finite $d$ ($d=2$ in most numerical simulations). For example, it predicts a diverging correlation length $\xi \sim \left| p-p_c\right|^{-4/3}$, with $p_c=1/2$ in this percolation limit. 

\subsubsection{Entanglement and minimal cut picture}

To compute the scaling of entanglement in this limit, it turns out to be more convenient to consider configurations with mixed measurement locations. Averaging over measurement outcomes and over Haar gates for a such given configuration of measurement locations, the statistical model mapping described above still goes through, with the link weight now being either $V_l(g_i^{-1} g_j) =d^{C(g_i^{-1} g_j}$ if that link is not measured, or $V_l(g_i^{-1} g_j) =d$ if that link coincides with a measurement. Now in the limit $d \gg 1$, we have  $V_l(g_i^{-1} g_j) \sim d^{Q} \delta_{g_i^{-1} g_j}$ as before so the statistical mechanics model for such a fixed measurement locations is a fully ordered (zero temperature) ferromagnet on a lattice diluted by the measurements:  each bond that is measured is effectively cut, while all other weights constrain the statistical model's spins to be the same in this limit. This is consistent with the percolation picture derived above.  

As we show next, a frustrated link  costs a large energy $\sim \log d$, leading to an effective minimal cut picture in that limit~\cite{Skinner2019}. To see this, recall that computing entanglement requires computing two different partition functions $Z_A$ and $Z_{0}$, which differs only by their boundary condition on the top boundary of the circuit. The boundary condition for the calculation of $Z_A$ forces a different boundary condition in region $A$, and thus introduces a domain wall (DW) near the top boundary. In the limit $d \to \infty$, the DW is forced to follow a minimal cut, defined as a path cutting a minimum number of unmeasured links (assumed to be unique for simplicity). Due to the uniform boundary condition in $Z_{0}$, all vertex elements in $Z_{0}$ are equal, so $Z_{0}$ is trivial and give by a single configuration of spins. $Z_A$ differs from $Z_{0}$ due to the fact the DW will lead to frustrated links that contribute different weights to $Z_A$. Each frustrated unmeasured link contributes a very large energy cost associated with this domain wall between $g=g_{\rm SWAP}$ and $g=e$
\begin{equation} 
\Delta E = (n-1) m \log d,
\end{equation}
using the energy weight on each link $E_l= - \log d \ C(g) $. Since this energy cost is very large as $d \gg1$, the domain wall will follow a path through the circuit minimizing the number of unmeasured links it has to cut. This leads to the expression $Z_A = p^{(1-n)m \ell_{\rm DW}}Z_{0}$, with $\ell_{\rm DW}$ the number of unmeasured links that the DW crosses along the minimal cut~\cite{Skinner2019,2021arXiv210710279A}. In the replica limit, this leads to a simple expression for the Renyi entropies 
\begin{equation} 
S^{(n)}_A =  \ell_{\rm DW}  \log d,
\end{equation}
where this equation is valid for any given configuration of measurement locations. We will use $\overline{ \ell_{\rm DW}} $ to denote the average of $\ell_{\rm DW}$ over measurement locations, which are simply percolation configurations. This quantity has a simple scaling in percolation: it is extensive  $\overline{ \ell_{\rm DW}}   \sim L_A$ (volume law) for $p_0<p_{0,c}=1/2$, and constant $\overline{ \ell_{\rm DW}}  \sim O(1)$ (area law) for $p_0>p_{0,c}=1/2$.
At criticality, this implies a logarithmic scaling of the entanglement entropy~\cite{1986JSP....45..933C, yao1612firstpassage, Skinner2019}
\begin{equation} \label{eqMinCut}
\overline{S^{(n)}_A}  \underset{d \gg 1}{\approx} \frac{\sqrt{3}}{\pi} \log d \log L_A.
\end{equation}
Strictly speaking, this minimal cut formula only applies for $d=\infty$, while for $d$ large but finite, it is only valid up for distances smaller than a crossover length $L_A \ll \xi(d)$ that we briefly discuss in the next section.  

\subsection{Finite $d$ universality class}

The infinite onsite Hilbert space dimension limit discussed above has an accidentally enlarged symmetry. The Potts model has a  symmetry group $S_{Q!}$ that is much larger than the $S_Q \times S_Q$ symmetry of the generic Boltzmann weights at finite $d$. (Note that $S_Q \times S_Q$ is a subgroup of $S_{Q!}$: the left and right action of $S_Q$ on itself has a permutation $g \in S_{Q!}$ representation -- this is known as Cayley's theorem.) The leading perturbation implementing this symmetry breaking in the Potts model was identified in Refs.~\onlinecite{PhysRevB.100.134203,Jian2020}, and turns out to be relevant, with scaling dimension $\Delta = \frac{5}{4}$. For any large but finite onsite Hilbert space dimension $d$, we thus expect a crossover from percolation criticality for length scales $\ell \ll \xi(d) \sim d^{4/3}$~\cite{Skinner2019}, to the finite $d$ universality class (which does not depend on $d$) at long distances $\ell \gg \xi(d)$. The boundary and bulk conformal spectrum of this theory were recently studied numerically~\cite{Li2020,Zabalo2020,2021arXiv210703393Z}, and  a Landau-Ginzburg action was proposed in Ref.~\cite{Nahum2020}. 
In the case of Clifford circuits, the universality class of the transition depends in a more subtle way on the onsite Hilbert space dimension~\cite{2021arXiv210712376Y,2021arXiv211002988L}.

\section{Symmetry and topology in measurement induced phases and criticality}
\label{SecSymmetryTopology}
Looking beyond featureless Haar-random two-qubit gates and single-site measurements, there is a huge variety of possible variations on a theme, including considering circuits and measurements that obey symmetry constraints, including multi-site measurements that ``collapse" into interesting entangled states rather than featureless product states, and many others. Here, we briefly describe a small selection of these enrichments to give a flavor of the possibilities.

\subsection{Symmetric monitored random circuits}
A simple extension of the MRC models described above is to include symmetry constraints in the random gates and measurement operations, i.e. demand that these preserve a symmetry group $G$. The appropriate symmetry depends on the microscopic realization of the qubits in question. For example, trapped ion and Rydberg atom systems natural have interactions that respect a discrete Ising symmetries ($G=\mathbb{Z}_2$), superconducting qubit and cavity-QED systems typically have $U(1)$-conserving interactions, and quantum dot spin-qubits often interact by $SU(2)$-invariant spin-exchange (absent spin-orbit coupling). Of these, so only simpler Abelian symmetries such as $G=\mathbb{Z}_2$~\cite{PhysRevResearch.2.023288,Sang2020,Lavasani2020,2021arXiv210804274L,2021arXiv210209164B} and $G=U(1)$~\cite{2021arXiv210208381B,2021arXiv210710279A} have been studied theoretically.

The inclusion of symmetry naturally begs two questions: i) are the universal properties of the entanglement transition modified by the symmetry? and ii) are there additional measurement-induced phases or critical phenomena that arise with symmetry [analogous to how symmetry distinguishes spontaneous-symmetry-broken and symmetry-protected topological (SPT) or symmetry-enriched topological (SET) phases in equilibrium]?
So far, the answer to the question i) appears to be negative~\cite{2021arXiv210710279A}, at least in the limit of large onsite dimension $d$ and small-scale numerics for $d=2$, it appears that the entanglement transition occurs in a regime where the charge degrees of freedom are frozen by measurements and do not affect the entanglement transition bulk criticality.
However, the answer to question ii) is affirmative, and numerous examples of transitions both in the area-law and volume- law phases have been constructed in a wide range of universality classes. 

To set the stage let us consider the general symmetry structure of the statistical mechanics replica models.
As detailed in~\cite{2021arXiv210209164B} (see also~\cite{Nahum2020}), in replicated statistical mechanics models a symmetry group $G$ of the circuit dynamics is incorporated into the $Q$-fold replicated theory as a separate $G$ symmetry separately for each replica and, within each replica, separately for both each forward ($U$) and backward ($U^*$) ``contours" of the time-evolution. These symmetry-factors are respectively permuted by the left- (forward time contour) and right (backward time contour) replica-permutation symmetry of the permutation ``spins" in the stat mech description. 
Finally, hermiticity of the density matrix implies that exchanging the forward and backward contours and complex conjugating the coefficients is a symmetry. Since doing this hermitian conjugation twice is trivial, this gives an extra $\mathbb{Z}_2$ factor to the symmetry group, but which acts non-trivially on the other replicated symmetry groups~\cite{Nahum2020}.
Combined, this gives overall symmetry group:
\begin{align}
\mathcal{G} = \[\(G_L^{\times Q}\rtimes S_{Q,L}\)\times \(G_R^{\times Q}\rtimes S_{Q,R}\)\]\rtimes \mathbb{Z}_2^{\mathbb{H}}
\end{align}
where $\rtimes$ indicates that the replica permutation action of the symmetric-group $S_Q$ also permutes the associated $G$ symmetries for each replica and hence these two factors do not generally commute, the left (L) and right (R) subscripts refer to the forward and backward contours respectively, and the $\mathbb{H}$ superscript on the final $\mathbb{Z}_2$ factor reminds that this is associated with hermiticity.
It is not yet clear whether or how the $\mathbb{Z}_2^{\mathbb{H}}$ factor plays a role in determining the structure of MIPTs, so we ignore it in the following, but simply mention it for completeness.

\subsection{Area-law phases}
As described above, the area- to volume- law entangled transition is a transition between phases in which the replica permutation symmetry respectively remains intact or becomes spontaneously broken. 

\begin{figure}[tb!] 
    \centering
    \includegraphics[width=1.0\textwidth]{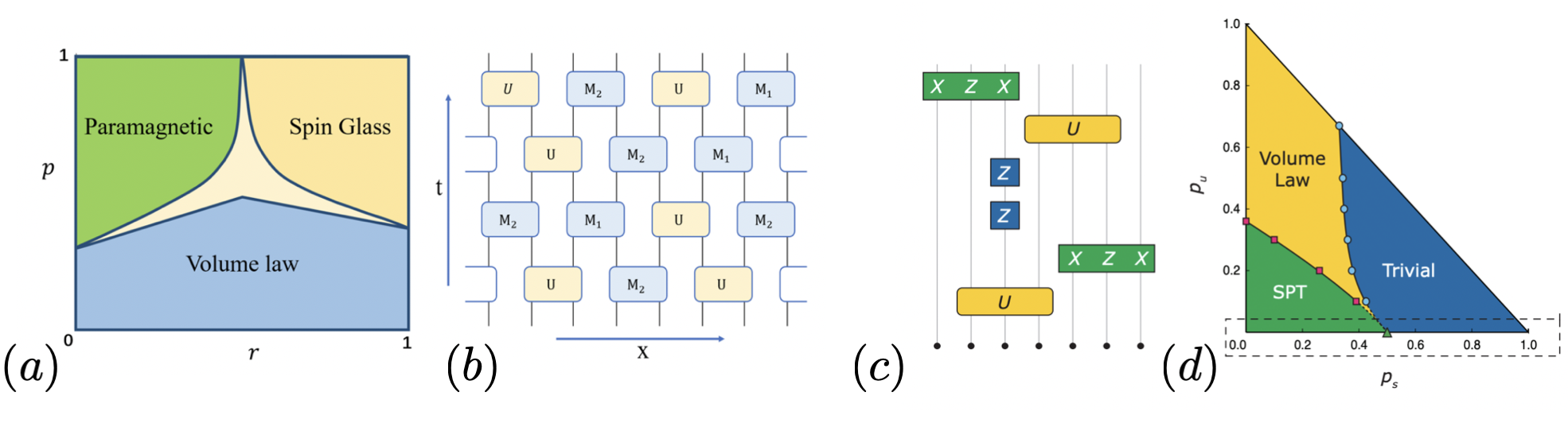}
    \caption{ {\bf Measurement induced stabilizer orders.} Phase diagram (a) and circuit model (b) of Ising-symmetric MRCs exhibiting area-law phases with- and without- order reproduced from Ref.~\cite{Sang2020}. The model in (a) consists of measurements with probability $p$ or random two-qubit clifford gates with probability $1-p$. Measurements are randomly chosen as $Z\otimes Z$ measurements with probality $r$ or single-qubit $X\otimes \mathbbm{1}$ or $\mathbbm{1}\otimes Z$ measurements with probability $1-r$. The phase diagram in (b) includes a critical fan (center region) which would shrink to a phase-boundary-line for an infinite system. (c-d) are reproduced from Ref.~\cite{Lavasani2020}, and show a model (c) that includes measurements of either $X\otimes Z\otimes X$ stabilizers for a cluster state (a $1d$ SPT protected by a pair of $\mathbb{Z}_2$ symmetries generated by $\prod_i Z_{2i}$ and $\prod_i Z_{2i+1}$ respectively), single-site $Z$ operators, and random clifford gates with probabilities $p_t,p_s, p_u$ respectively. In the measurement dominated regime (dashed box), $p_s$ tunes between area-law phases with trivial or SPT order. These give way to volume-law entangled phases at sufficiently large $p_u$.
    \label{fig:u1} 
            }
\end{figure}
Since the short-range entanglement structure of the area-law phase is identical to that found in ground-states of local Hamiltonians, it is natural to guess that the area-law MRC phases with symmetry $G$ coincide with equilibrium phases (i.e. paramagnetic, spontaneous symmetry-broken, SPT, or SET) with the same symmetry. 

Indeed, for discrete symmetries it is straightforward to design models that achieve a large class of symmetry-breaking and topological orders in the measurement-dominated regime. To be specific, consider Ising symmetric random Clifford circuits ($G=\mathbb{Z}_2$), with (spacetime-random) measurements drawn from local generators of a stabilizer group $\mathcal{S}$ (i.e. a group of mutually-commuting Pauli strings). In the extreme limit of measurement-only dynamics, these stabilizer measurements project into a state specified by eigenvalues of $s=\pm 1$ for each $s\in \mathcal{S}$. Such stabilizer states can support a large class of interesting many-body orders including discrete symmetry breaking and symmetry-protected topological (SPT) orders, and non-chiral- topological- or fracton- orders (i.e. toric code like orders but not fractional quantum Hall effect with chiral edge modes). 

\subsubsection{Measurement-induced symmetry-breaking order in $1+1d$}
For example, in a qubit chain with symmetry generated by $X=\prod_i X_i$, measuring $s\in \{Z_iZ_{i+1}\}$ on every bond projects into random spin-glass state(s) with frozen but random spin-texture $s_{i,i+1}=Z_iZ_{i+1}=\pm 1 \forall i$. Here, a fixed set of measurement outcomes $s_i$, actually correspond to two possible states. E.g. for a 3-site chain states with $\{s_{12},s_{23}\} = \{+1,-1\}$ form a two-dimensional subspace with basis states $\{|\up\up\down\>,|\down\down\up\>\}$ that each spontaneously break the $\mathbb{Z}_2$ symmetry (or equivalently we can form cat-like superpositions of $|\up\up\down\>\pm |\down\down\up\>$ which have definite overall $X=\pm 1$, but have long-range mutual information between all spins). Thus, we see that the states stabilized by this measurement-only dynamics have the same form as the spontaneous symmetry breaking ground-space of an ideal Ising magnet with couplings that are ferromagnetic or antiferromagnetic depending on $s_{i,i+1}=\pm 1$. 

We note that, different trajectories have different random frozen configurations, so that the trajectory average of long-range symmetry breaking correlations such as $\lim_{|i-j|\rightarrow \infty} \mathbb{E}_{m,U}\<Z_iZ_j\> = 0$ strictly vanish (this could be anticipated on general grounds since such linear averages always behave like infinite temperature averages), but where higher-moments of symmetry-breaking correlations such as the Edwards-Anderson (EA) type order parameter $\chi^{(2)} = \lim_{|i-j|\rightarrow \infty} \mathbb{E}_{m,U}|\<Z_iZ_j\>|^2$ are non-vanishing.

So far, we have considered an idealized limit with only measurements. Numerical simulations~\cite{Sang2020,2021arXiv210209164B,Lavasani2020} that perturb away from this fine-tuned point by including random unitary dynamics (e.g. by random Clifford circuits, which can be efficiently simulated) or by competing stabilizer measurements (e.g. of non-commuting observables such as $\{X_i\}$ that stabilize trivial symmetric product states), show clear evidence that these area-law orders survive over a finite range of couplings in thermodynamically large systems, and extend to generic area-law phase with symmetry breaking order in trajectories with area-law entanglement. We note that, generically, the spin-glass pattern found in the ordered trajectories is not frozen in time, but undergoes (classical) stochastic fluctuations induced by competing measurements or unitary gates between each time-step. 

\subsubsection{Measurement-induced topological orders}
The above recipe can be straightforwardly extended to produce models of area-law phases with more complicated measurement-induced orders including: SPT orders~\cite{Lavasani2020}, intrinsic topological or SET order, and fracton orders -- essentially any phase describable by stabilizer states. As for the measurement-induced spontaneous symmetry breaking phase discussed above, these phases are characterized by glassy order in each trajectory, which can be diagnosed by various non-local analogs of the EA order parameter described above. Due to its relation to error-correction, we briefly elaborate on a particular example: measurement induced $\mathbb{Z}_2$ (a.k.a. ``toric"- or ``surface"- code) topological order~\cite{maissam} in $2+1d$ MRCs.

Following~\cite{maissam}, consider a standard model of the surface code consisting of a qubit arrays on a chess board (square lattice with two-site unit cell consisting of alternating ``black" and ``white" squares) , whose dynamics are dominated by measurements of stabilizers consisting of product of $X$ or $Z$ operators respectively over the corners of the white and black squares or ``plaquettes" (P), $s_P \in \{\prod_{i\in P_\text{white}} X_i,\prod_{i\in P_\text{black}} Z_i\}$. In the measurement-dominated regime, measurements then collapse the state into one adiabatically connected to a stabilizer state $s_P \in \{\pm 1\}$.  If this model is defined on closed manifold, then specifying $s_P$ for all plaquettes, $P$ uniquely specifies a state. However, if the model is defined on a topologically non-trivial manifold, $M$, with genus $g$, e.g. a torus ($g=2$), then the stabilizers only fix a ground-space of dimension $4^g$, which can be seen by noting that for each nontrivial cycle $c\in \pi_1(M)$, around a handle of $M$, one can define a pair of additional stabilizers, $X_c,Z_c = \prod_{i\in M} X_i,Z_i$ which are independent from, but commute with the measured stabilizers $s$. $X_c$ and $Z_{c'}$ operator loops that intersect but wrap different cycles anticommute, and hence cannot have simultaneous eigenstates. These topological ground-space has been proposed as a promising quantum memory for fault tolerant computation, since its states cannot be distinguished by local noise, but rather, only by measuring non-local string operators $(X,Z)_c$.

This gives rise to the topological degeneracy in the measurement-stabilized trajectories. In Ref.~\cite{maissam}, it was shown that by tuning the fraction of surface-code stabilizer measurements with competing random circuit evolution or trivial product-state stabilizer measurements, one can obtain phase transitions between trajectories with- and without $\mathbb{Z}_2$ topological order. Unlike for the symmetry-breaking example above, this distinction cannot be probed by any local measurement. Instead, one can diagnose topological order either by two alternative routes. Each trajectory would exhibit a quantized entanglement entropy $\gamma=\log 2$~\cite{Kitaev,PhysRevLett.96.110405,maissam} defined via the subleading constant part of the entanglement of a region $A$: $S_A = a|\partial A|-\gamma$ where $a$ is a non-universal constant, and $|\d A|$ is the length of the perimeter of region $A$. 

Alternatively, one could consider the dynamics of a maximally mixed state: here, in the area-law phase, the measurements tend to purify the state. However, in the topologically ordered phase, measurements only purify the local degrees of freedom, and leave a equal-weighted mixture over the $4^g$ states of the topological ground-space resulting in entanglement $S\sim 2g\log 2$ in the steady state.

The model of Ref.~\cite{maissam} also exhibits an unconventional critical point with non-relativistic dynamics (dynamical exponent $z\neq 1$), and logarithmic violations of area-law scaling in $2d$, possibly related to emergent subsystem symmetries. This highlights the potential for novel types of critical phenomena arising in MRC dynamics, which do not naturally arise in more familiar equilibrium settings.

\subsection{Volume-Law phases}
The prospect of measurement-induced orders in the area-law phase could be anticipated based on our knowledge of ground-state phases of equilibrium matter, which also exhibit area-law entanglement.
Perhaps more surprising is the prospect of additional phases arising within the volume law entangled regime, where the entanglement structure of trajectories is similar to that of finite-temperature equilibrium states, which do not support any quantum orders and rule out even classical orders in low dimensions. 

\subsubsection{Volume-Law phases with order -- stat mech perspective}
In the statistical mechanics language, the volume-law regime corresponds to spontaneously broken $S_{Q,L}\times S_{Q,R}$ replica permutation symmetry. However, the residual $G_{L,R}^{\times Q}$ symmetries remain and could conceivably be spontaneously broken, or protect- or enrich- topology.
These various possibilities are discussed in detail in Ref.~\cite{2021arXiv210209164B}, which provided a particularly simple argument (supported by numerics) for volume-law phases with order: Consider stacking (and weakly coupling) two subsystems that, before coupling, respectively form an ordered area-law phase, and a featureless volume-law phase of degrees of freedom with trivial symmetry properties. Since, each of these systems corresponds to gapped phases in the stat-mech variables, weak coupling between the two will not destroy the symmetry-breaking or topological properties of the first subsystem, nor the volume-law entangled (replica-permutation breaking order) of the second subsystem, and will therefore result in an ordered and volume-law entangled phase.
While ``obvious" in the stat-mech language, this predicts a highly-nontrivial result in terms of the original degrees of freedom in the MRC. Namely, that it is possible to have stable phases with quantum-coherent orders in the highly-entangled and scrambled trajectories of a quantum circuit! The coexistence of volume-law entanglement and symmetry-breaking spin glass order (below the classical lower critical dimension) was also observed numerically in symmetric Clifford circuits in Refs.~\cite{Sang2020, 2021arXiv210804274L}.

\subsubsection{Charge sharpening transitions in the volume law phase}
Beyond the possibility of stabilizing ground-state orders in volume-law entangled trajectories, it turns out that there are additional phases and associated critical phenomena within the volume-law regime of symmetric MRCs that cannot be understood by any ground-state order parameter, but rather are distinguished by sharply distinct dynamics of symmetry quantum numbers (which we henceforth refer to as ``charges"). 

\begin{figure}[tb!] 
    \centering
    \includegraphics[width=0.5\textwidth]{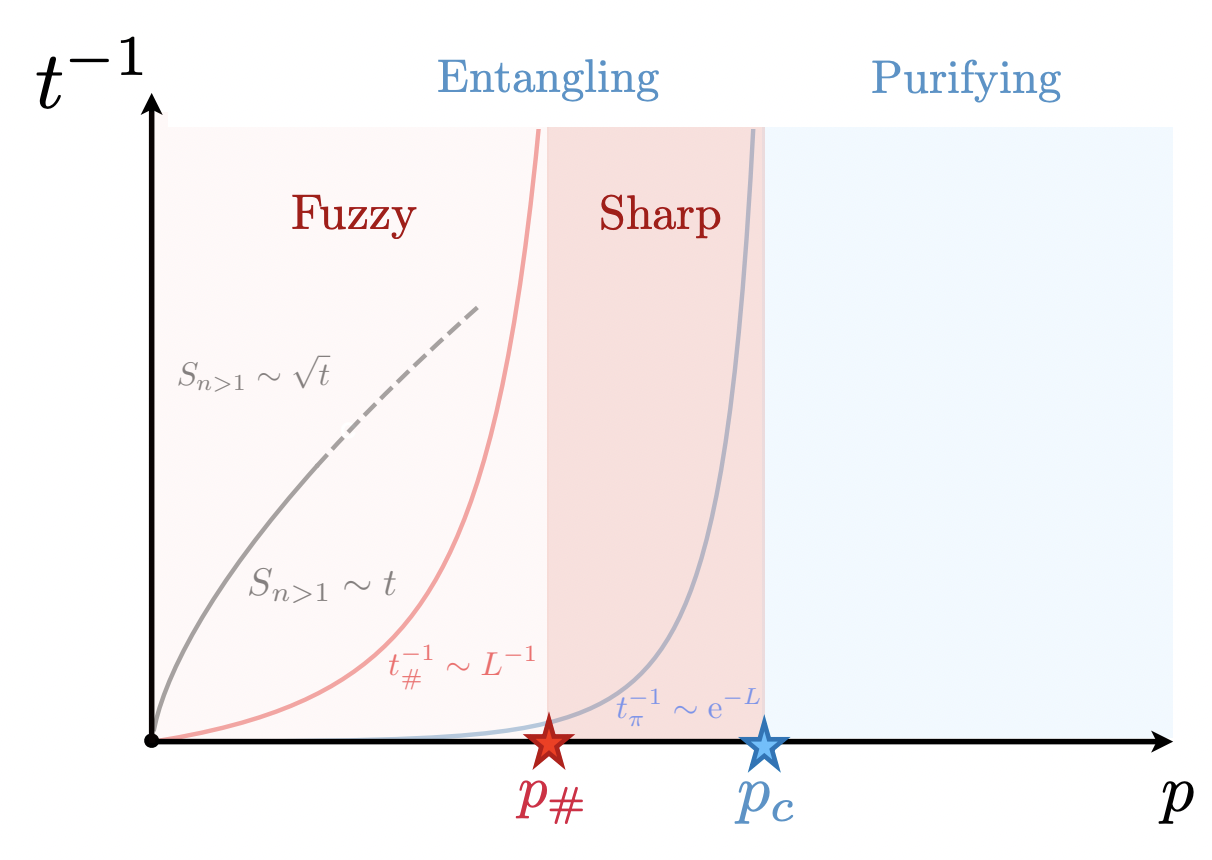}
    \caption{ {\bf Schematic phase diagram of $1+1d$ U(1) Symmetry MRCs} for the $U(1)$ symmetric MRC model described in the text (reproduced from Ref.~\cite{2021arXiv210710279A}), includes both a volume-law entanglement (red) to area-law entanglement (blue) transition at critical measurement rate $p=p_c$, and a charge sharpening transition within the volume law phase at $p=p_\#$. The diffusive growth of Renyi entropies, $S_n$, associated with diffusive dynamics and rare dead-region physics, for random unitary circuit evolution $p=0$, immediately converts to ballistic growth for any non-zero measurement rate $p>0$. $t_\#$ denotes the time scale for measurements to sharpen an initial state that is a superposition of different charge sectors into one with definite charge. The purification time for an initially mixed state to collapse into a pure state due to measurements is denoted by $t_\pi$. $L$ denotes system size.
    \label{fig:u1} 
            }
\end{figure}

For concreteness, let us consider augmenting the $U(1)$-symmetric RC model described above in section~\ref{SectionU1}, by adding random measurements of each site with probability $p$ in the charge-basis of the qubit and in any basis of the neutral large-$d$ qudit~\cite{2021arXiv210710279A}. 
This model can be analyzed by generalizing the statistical mechanics model to incorporate symmetry, which results in hardcore charge degrees of freedom that undergo a random walk on the replicated circuit network, and which are coupled to the replica-permutation spins by the measurements~\cite{2021arXiv210710279A}. In the large-$d$ limit, it is possible to take the replica limit exactly, and analyze the charge dynamics as a classical stochastic evolution.
In the large-$d$ limit, standard Haar averaging formulas immediately imply that off-diagonal coherences between density matrix elements of different charge are strictly vanishing, i.e. this model does not support any spontaneous breaking of $U(1)$ symmetry. Nevertheless, two distinct phases are observed~\cite{2021arXiv210710279A} within the volume-law entangled regime, separated by separated by an apparently continuous phase transition at critical measurement probability $p_\#<p_c$ which precedes the entanglement transition at $p_c$ ($p_c=1/2$ in the large-$d$ limit). 

For $p<p_\#$, the measurements fail to collapse an initial superposition of different total charges into a given charge sector, i.e. the global symmetry quantum number (``charge") remains ``fuzzy" up to a time-scale $t\gtrsim L$ which diverges with system size $L$. By contrast, for $p>p_\#$, measurements collapse the system into a state with sharp total charge in finite time independent of system size (for $t\gg L$ measurements always sharpen the total charge). In $1+1d$ models, these charge-``sharp" and charge-``fuzzy" phases are distinguished by the behavior of charge fluctuations~\cite{2021arXiv210710279A,2021arXiv211109336B}:
\begin{align}
C(r)=\mathbb{E}_{m,U}\[\<\sigma^z(r)\sigma^z(0)\>-\<\sigma^z(r)\>\<\sigma^z(0)\>\] \sim
\begin{cases}
1/r^2 & p\leq p_\# \\
e^{-r/\xi} & p>p_\#
\end{cases},
\end{align}
where the power-law decay for $p<p_\#$ is expected to become truly long-ranged $\lim_{r\rightarrow\infty}C(r)\sim \text{const.}$ in higher dimensions, or with discrete Abelian symmetries~\cite{2021arXiv211109336B}. In addition, this work reveals that even a small amount of measurements singularly change the dynamics of entanglement growth and charge fluctuations from the diffusive (dynamical exponent $z=2$) motion for $p=0$, to ballistic ($z=1$) for $0<p\leq p_\#$, and eventually to exponentially fast relaxation ($z\rightarrow\infty$, massive dynamics) for $p>p_\#$. 

We emphasize that, unlike the measurement-induced symmetry-breaking and topological- orders described above, which are smoothly connected to equilibrium ground-state orders\footnote{e.g. by continuously turning off the coupling between the stabilizer-state qubits and volume-law entangled trivial degrees of freedom in the above construction, and then dialing the stabilizer measurement probability to unity, which, in the replicated statistical mechanics description corresponds to disentangling two gapped degrees of freedom and then smoothly changing couplings within a gapped phase respectively, and does not produce a phase transition.} , this charge-sharpening transition (like the entanglement transition) is a purely dynamical effect that is special to non-equilibrium MRC dynamics. While these examples give an idea of the possibilities for new types of non-equilibrium measurement-induced orders, at present, a rigorous/exhaustive classification of measurement-induced orders remains an open challenge.

\section{Discussion}
The examples reviewed above highlight the promise for using well-developed statistical mechanics tools to investigate universal aspects of emerging quantum dynamics and quantum information theory concepts, and to uncover new regimes of measurement-induced non-equilibrium orders.

Despite rapid progress, several open challenges remain. The statistical mechanics model mapping of MRC dynamics onto a classical spin model establishes the existence of these transitions through convergence of strong- and weak- coupling expansions, establishes an equivalence between the entanglement- and purification- transition perspectives, and strongly suggests that the $1+1d$ entanglement transition is described by a (non-unitary) conformal field theory (CFT). However, a detailed analytic understanding of the precise universality class and CFT content remains unsolved (though these questions are being explored numerically~\cite{Li2020,Zabalo2020,2021arXiv210703393Z}). To gain analytic insight into this question, and particularly to study measurement induced criticality and orders in $2d$ and $3d$ circuits and incorporate more complex features such as multiple types of non-commuting measurements, it may be desirable to develop a continuum quantum field theory methods for treating quantum circuits (see \cite{2021arXiv210208381B} for preliminary efforts along these lines for free-fermion circuits). The study of different classes of circuits, including {\it e.g.} Gaussian free-fermion circuits and tensor networks~\cite{10.21468/SciPostPhys.7.2.024,Chen2020,2020arXiv201204666J,2021arXiv210208381B,PhysRevLett.126.170602,2021arXiv211103500T}, or self-dual unitary circuits~\cite{PhysRevLett.123.210601,PhysRevB.100.064309}, might also help addressing some of those questions, as well as revealing new measurement-stabilized phases.
Looking beyond these more detailed aspects, a natural question is whether there are other paradigms for obtaining universal phenomena in quantum circuit dynamics, for example, which do not require post-selection on measurement outcomes to observe and could be explored experimentally as well as theoretically.

Early studies have also suggested intriguing possibility of gaining universal insights into fundamental limits of quantum communication. However, there is significant room to put these suggested connections on more rigorous footing, and to explore other connections between statistical mechanics of random circuits to random quantum error-correcting codes, or possibly even to fundamental quantum complexity theory. One could even imagine that studies of noise and error propagation in random circuits might yield practical design principles for optimizing aspects (e.g. dimensionality, connectivity, etc...) of qubit architectures in order to maximize their ability to generate complicated entangled states or minimize the impact of noise on quantum algorithms.

\section*{Acknowledgments}
We thank our collaborators Utkarsh Agrawal, Fergus Barratt, Matthew Fisher, Aaron Friedman, Sarang Gopalakrishnan, Michael Gullans, David Huse, Chao-Ming Jian, Yaodong Li, Andreas Ludwig, Adam Nahum,  Javier Lopez-Piqueres, Jed Pixley, Hans Singh, Yi-Zhuang You, Brayden Ware, Justin Wilson, and Aidan Zabalo for many insightful discussions. We also thank Ehud Altman, Maissam Barkeshli, Xiao Chen, Michael Gullans, Tim Hsieh, Yaodong Li, Adam Nahum and Jed Pixley for helpful comments on this manuscript.
 This research was supported in part from the US Department of Energy, Office of Science, Basic Energy Sciences, under Early Career Award No. DE-SC0019168 (RV), from the US National Science Foundation DMR-1653007 (ACP), and the Alfred P. Sloan Foundation through Sloan Research Fellowships (RV and ACP). This research was undertaken thanks, in part, to funding from the Max Planck-UBC-UTokyo Center for Quantum Materials and the Canada First Research Excellence Fund, Quantum Materials and Future Technologies Program (ACP).

\appendix

\bibliography{references}
\bibliographystyle{apsrev4-1}

\end{document}